\documentclass[journal]{IEEEtran}
\usepackage{cite}
\ifCLASSINFOpdf
\else
\fi
\usepackage{adjustbox}

\usepackage{amsmath,amssymb,amsthm}
\usepackage[algoruled, linesnumbered]{algorithm2e}

\SetCommentSty{mycommfont}

\ifCLASSOPTIONcompsoc
  \usepackage[caption=false,font=normalsize,labelfont=sf,textfont=sf]{subfig}
\else
  \usepackage[caption=false,font=footnotesize]{subfig}
\fi

\usepackage{booktabs}
\usepackage{multirow}

\usepackage{url}
\usepackage{hyperref}

\usepackage[utf8]{inputenc}
\usepackage[usenames, dvipsnames]{color}
\usepackage{xcolor}

\usepackage{tikz, pgfplots}
\pgfplotsset{compat=newest}
\usetikzlibrary{plotmarks}
\tikzset{every picture/.style=thick}
\pgfplotsset{
	compat=newest,
	minor grid style={dotted},
	every axis/.append style={
		title style={font=\bfseries}
	}
}
\pgfkeys{/pgf/number format/.cd, 1000 sep={\,}} % tisícový oddělovač mezera

\usepackage[textsize=scriptsize, textwidth=1.35cm, colorinlistoftodos]{todonotes}
\def\RR{\mathbb{R}}

\newcommand{\vect}[1]{\mathbf{#1}}
\def\x{\vect{x}}
\def\uu{\vect{u}}
\def\vv{\vect{v}}
\def\y{\vect{y}}

\def\a{\vect{a}}

\def\e{\vect{e}}

\newcommand{\matr}[1]{\mathbf{#1}}
\def\X{\matr{X}}
\def\A{\matr{A}}
\def\M{\matr{M}}
\def\Id{\matr{I}}

\def\Mr{\M_\textup{R}}
\def\Mm{\overline{\M}_\textup{R}}
\def\Mh{\M_\textup{H}}
\def\Ml{\M_\textup{L}}
\def\tc{\theta_\textup{clip}}

\def\theset{\varGamma}
\def\setinp{\theset_\textup{inp}}
\def\setdec{\theset_\textup{declip}}
\def\setdeq{\theset_\textup{dequant}}

\def\fcoef{f_\textup{C}}
\def\fsig{f_\textup{S}}
\def\lcoef{\lambda_\textup{C}}
\def\lsig{\lambda_\textup{S}}
\def\gcoef{\gamma_\textup{C}}
\def\gsig{\gamma_\textup{S}}
\def\tcoef{\tau_\textup{C}}
\def\tsig{\tau_\textup{S}}

\newcommand{\norm}[1]{\|#1\|}
\newcommand{\abs}[1]{\left\vert#1\right\vert}
\newcommand{\transp}[1]{#1^\top}

\newcommand{\argmin}{\mathop{\operatorname{arg\,min}}}

\newcommand{\prox}{\mathrm{prox}}

\newcommand{\proj}{\mathrm{proj}}

\newcommand{\sgn}{\mathrm{sgn}}

\newcommand{\edt}[1]{{#1}}

\newcommand{\qm}[1]{``#1''}

\begin{document}
	
%\listoftodos
%\clearpage

% \title{Regularized autoregressive modeling \\ and its application to audio signal declipping}
%	\ond{restoration/reconstruction/recovery/declipping}}
\title{Regularized autoregressive modeling \\[-2mm] and its application to audio signal reconstruction}

\author{%
	Ondřej Mokrý,
	Pavel Rajmic%
	%, Pavel Záviška% <-this % stops a space
\thanks{%
	O. Mokrý and
	P. Rajmic % and P. Záviška
	are with the SPLab at the Faculty of Electrical Engineering and Communication, Brno University of Technology, Czech Republic. E-mail:
\href{mailto:ondrej.mokry@vut.cz}{\texttt{ondrej.mokry@vut.cz}},
\href{mailto:pavel.rajmic@vut.cz}{\texttt{pavel.rajmic@vut.cz}},
%\href{mailto:xzavis01@vutbr.cz}{\texttt{xzavis01@vutbr.cz}}
}}


\markboth{IEEE Transactions on Audio, Speech, and Language Processing}%
{Mokrý \MakeLowercase{\textit{et al.}}: Regularized autoregressive modeling}



\maketitle

\begin{abstract}
%	* AR modeling in signal processing is invaluable, in particular in speech and audio.
%	* There are attempts in the literature to regularize or constrain either the time-domain signal values or the AR coefficients, which is done for various reasons, including the incorporation of a~prior information or numerical stabilization.
%	* Although these represent appealing strategies, an encompassing and general enough modeling framework is still missing.
%	* We present such a framework and the related optimization problem and algorithm.
%	* We discuss also its computational demands and the effects of various improvements on its convergence.
%	* We demonstrate its usefulness on the audio declipping problem.
%	* We compare its performance against state-of-the-art methods and show that we are on par with them, especially for mildly clipped signals.
%	* The heuristic generalized linear prediction (GLP) algorithm, so far presented only in the form of a patent, is also included in the evaluation.
%
Autoregressive (AR) modeling is invaluable in signal processing, in particular in speech and audio fields.
Attempts in the literature can be found that regularize or constrain either the time-domain signal values or the AR coefficients, which is done for various reasons, including the incorporation of prior information or numerical stabilization.
Although these attempts are appealing, an encompassing and generic
% general-enough
modeling framework is still missing.
We propose such a framework and the related optimization problem and algorithm.
We discuss the computational demands of the algorithm and explore the effects of various improvements on its convergence speed.
In the experimental part, we demonstrate the usefulness of our approach on the audio declipping and dequantization problems.
We compare its performance against state-of-the-art methods and demonstrate
% that \pav{we are on par with them --- to se mi nelíbí ale nevím co s tím},
the competitiveness of the proposed method in declipping musical signals,
%especially for mildly clipped signals.
and its superiority in declipping speech.
%\todo{PR: in certain setups; respektive tuto druhou část věty bych ještě přizpůsobil podle výskedků i na řeči}
The evaluation includes a~heuristic algorithm of generalized linear prediction (GLP),
a~strong competitor which has only been presented as a~patent and is new in the scientific community.
% The heuristic generalized linear prediction (GLP) algorithm, so far presented only as a~patent, is also included in the evaluation.
\end{abstract}

\begin{IEEEkeywords}
autoregression, regularization, inverse problems, audio declipping, proximal splitting methods, sparsity
\end{IEEEkeywords}

% \fi
\IEEEpeerreviewmaketitle

\section{Introduction}
\IEEEPARstart{A}{utoregressive} (AR) models have been used in a~variety of
%signal-processing
engineering problems.
Tasks currently solved with the help of AR modeling include speech coding
%použitelné jako model řeči/audiosignálu pro analýzu a kódování -- linear predictive coding
\cite[Sec.\,8.2.2]{Zolzer2011:DAFX},
music analysis and coding
\cite[Ch.\,4]{AudioSignalProcessingAndCoding},
and time series prediction in finance \cite[Ch.\,2]{Tsay2005:Analysis.financial.time.series},
for instance.
AR processes have found their use also within control theory
\cite{Akaike1998:Autoregressive.Model.Fitting.for.Control},
where a~clear connection to Hankel operators exists%
%is most prominent
\cite{Takahashi2013:Hankel_matrix_declipping,Zaviska2023:Multiple.Hankel.matrix.inpainting}.

Fitting a~signal with an AR process possesses a~kind of inverse problem.
Thus, where appropriate, it has found its use in reconstruction tasks where data are missing or where data are observed in a~degraded form.
In its simplest form, the AR fitting of a signal can be understood as a~denoiser---the noise is identified with the residual error.
%which can be obtained by inverse filtering the noisy signal.
% , and then subtracted from it.
A~more complicated application example is
%More advanced applications include Further simple applications include 
the time series extrapolation or interpolation of missing segments of audio \cite{Etter1996:Interpolation_AR}.
In some inverse problems, the prior knowledge about the signal can consist of more than only its AR nature.
In such advanced cases, however, a~need often arises to regularize or constrain the signal itself or even the AR coefficients.
Common AR fitting algorithms are not suitable to cope with such augmented situations.

The contribution of this article is twofold.
First, it presents general models which, beside common AR assumptions,
allow additional regularization terms, and offers related numerical algorithms.
Second, the theory is validated on audio declipping and dequantization problems.
%, and
%the proposed approach is compared with the state of the art methods.
%\todo{PR: ale dekvantizaci neporovnáváme}

%\todo[inline]{\textbf{PR:} Z Úvodu by mělo být zřejmé, že budeme dělat declipping.
%	Nemělo by to být zřejmé i z názvu?!}

\subsection{Audio inverse problems}
% Audio inpainting and audio declipping are two popular signal recovery tasks.
Audio inpainting, declipping and dequantization are popular signal recovery tasks.
In audio inpainting, signal samples
%or their blocks
are missing, and the inverse problem consists in reconstructing (i.e., estimating) them.
Within an optimization framework, which is the most widely used recovery approach, this can be formulated as 
\begin{equation}
	\label{eq:Inpainting.problem}
	\min_{\x} f(\x)\ \,\text{s.t.}\ \,\x \in \setinp = \{ \x\in\RR^N \!\mid\! \Mr\x = \Mr\y \}.
\end{equation}
In this problem, a~signal $\x$ is sought such that it meets the requirements expressed by the penalization function $f$.
At the same time, $\x$ must
%be identical to
perfectly fit
the observed portion of the signal $\y\in\RR^N$\!.
%, at the samples under observation.
%\todo{\textbf{OM:} Opraveno podle DK, ale přikláněl bych se za poslední čárkou smazat.}
% Here, the masking operator $\Mr$ identifies the reliable samples in question.
\edt{%
Here, the matrix $\Mr$ represents a mask of the reliable samples in question, effectively performing row selection.
}
The condition $\x \in \setinp$ is often called the consistency condition~\cite{Kitic2013:Consistent.iter.hard.thresholding}.

In audio declipping, samples are not completely missing, but, instead, they are observed saturated (clipped), i.e.,
samples exceeding the dynamic range $[-\tc, \tc]$ are limited in amplitude according to
the nonlinear formula
\cite{ZaviskaRajmicOzerovRencker2021:Declipping.Survey,ZaviskaRajmicMokry2022:Declipping.crossfading}
\begin{equation}
	y_n = \left\{
	\begin{aligned}
		&x_n &\text{for} \hspace{1em} &|x_n| < \tc, \\
		&\tc \cdot \sgn^{\!+\!}(x_n) &\text{for} \hspace{1em}  &|x_n| \geq \tc.
	\end{aligned}
	\right.
	\label{eq:clipping}
\end{equation}
Here, \edt{$\sgn^{\!+\!}(z)$ returns $1$ for $z\geq0$ and $-1$ for $z<0$ \cite{ZaviskaRajmicMokry2021:Audio.dequantization.ICASSP}},
and $\x$~is the input signal and $\y$ is its observed counterpart, see an example in Fig.\ \ref{fig:waveforms.consistency}.
%
%\todo[inline]{PRatOM: zde dát tikz obrázek podobný jako v jedné z citovaných publikací; mělo by tam být vidět konzistentní a nekonzistentní řešení}%
%
\begin{figure}[t]
	\centering
	\adjustbox{width=\linewidth}{\input{figures/waveforms_consistency.tex}}%
	\vspace{-3mm}%
	\caption{Illustration of audio clipping and of the consistency property of the declipping solutions; adapted from~\cite{ZaviskaRajmicMokry2022:Declipping.crossfading}.}
	\label{fig:waveforms.consistency}
\end{figure}
In such a~case, the recovery problem reads
\begin{equation}
	\label{eq:declipping.problem.generally}
	\min_{\x} f(\x)\ \text{subject to}\ \x \in \setdec,
\end{equation}
where the consistency set $\setdec$ corresponds to the element-wise clipping constraints $\pm\tc$:
\begin{equation}
	\label{eq:Gamma.declip.definition}
	\setdec =
	% \{ \x\in\RR^N \mid \Mr\x = \Mr\y, \Mh\x \geq \tc, \Ml\x\leq -\tc \}.
	\left\{ \x\in\RR^N \mid
	\begin{array}{l}
		\Mr\x = \Mr\y \\
		\Mh\x\geq \tc \\
		\Ml\x\leq -\tc 
	\end{array}
	\right\}.
\end{equation}
%
%The respective masking operators $\Mr$, $\Mh$ and $\Ml$ serve to pick samples from the 
%%corresponding
%set of indices $\{1,\dots,N\}$~\cite{ZaviskaRajmicMokry2022:Declipping.crossfading}.
\edt{%
The respective matrices $\Mr$, $\Mh$ and $\Ml$ serve 
to divide the vector $\x$ into the non-clipped part $\Mr\x$ and the samples exceeding $\tc$ and $-\tc$, respectively
\cite{ZaviskaRajmicMokry2022:Declipping.crossfading}.
}

In declipping, %on one hand, 
more information is available about the signal than in inpainting.
Unfortunately,
%But on the other hand,
while the restriction $\x\in\setinp$ in \eqref{eq:Inpainting.problem} still allows solving it via the use of common AR fitting algorithms,
the additional knowledge coded in $\x\in\setdec$ is not directly transferable into these algorithms, as will also be discussed in the following part.

As the last example,
quantization rounds each signal sample to the nearest from a limited number of quantization levels \cite{Zolzer2011:DAFX}.
Several ways to define the quantization levels are available~\cite{ITU-T_G.711};
in the standard case of the mid-riser uniform quantizer,
the quantization levels are equally spaced %distributed. % over the whole dynamic range.
and symmetric around zero.
For a~word length of $w$ bits per sample (bps), 
the quantization step is $\Delta = 2^{-w+1}$
%is constant in the whole dynamic range %PR: to už zaznělo výše
and the quantization process follows the equation
%\todo{Označit pozorovaný signál jinak, např. $\y^{\text{quant}}$?\\ PR: Výše jsme to taky nerozlišovali, nechal bych.}
\begin{equation}
	y_n = \sgn^{\!+\!}(x_n)\cdot \Delta \cdot \left( \left\lfloor \frac{|x_n|}{\Delta} \right\rfloor + \frac{1}{2} \right)\!.
	\label{eq:uniform_quantization}
\end{equation}
For the dequantization inverse problem,
\eqref{eq:uniform_quantization} induces the consistency set
\begin{equation}
	\setdeq = \{\x\in\RR^N \mid \norm{{\x - \y}}_\infty\! < \Delta/2\}.
	\label{eq:Gamma.dequant.definition}
\end{equation}
In comparison to both inpainting and declipping,
the constraint in recovering the degraded samples is the most narrow;
however,
this time no reliable samples are available.

%\todo{PR: Přidat zmínku a citaci našeho DAFx, že totéž může být navíc v nějaké transformované doméně? \\ *: Spíš ne, bylo by to trhání pozornosti.}

%\subsection{Audio declipping methods}
\subsection{Autoregressive methods for signal reconstruction}
\label{sec:AR.methods.for.signal.reconstruction}

It is clear that any inpainting method can be used for declipping as well
%This is achieved 
by simply ignoring the conditions involving $\Mh$ and $\Ml$ in \eqref{eq:Gamma.declip.definition}.
This is also the case of the Janssen method \cite{javevr86,Oudre2018:Janssen.implementation},
which is the state-of-the-art for audio inpainting
for compact signal gaps
%up to 50\,ms in length~\cite{MokryRajmic2020:Inpainting.revisited}.
of up to 80\,ms in length~\cite{MokryRajmic2020:Inpainting.revisited, MokryRajmic2025:Inpainting.AR}.
%\todo{Někde tady ocitovat \cite{MokryBalusikRajmic2025:Spectrogram.inpainting}, ať ho nemusíme zmiňovat v poznámkách k submission.\\
%*: Nedávejme ho sem, je to tématicky jiné.}
It~is a~blockwise, iterative approach where in each iteration,
the AR coefficients are estimated for a~whole block of a~signal and then the missing samples are updated according to the model.
In the next iteration, the model is re-estimated based on the current solution and the process is repeated in such a~manner.
Importantly, the iterations involve standard AR estimation only.

The patent \cite{AtlasClark2012:Generalized.linear.prediction}
proposes a~method for audio declipping,
referred to as generalized linear prediction (GLP),
that builds upon the idea of Janssen.
The extension lies in that after each iteration,
all samples currently violating the conditions
$\Mh\x \geq \tc, \Ml\x\leq -\tc$
are flipped around the respective $\pm\tc$ levels.
This is quite a~heuristic step but it surprisingly leads to an algorithm that is one of the best performing.
The flipping operation on the signal, described by the authors as a~rectification, forces the consistency in the clipped part.
Interestingly, the method
%(the same as the patent)
is practically ignored by the signal processing community.
%\todo{\textbf{PR:} a pravděpodobně jsme první, kdo ji zmiňuje v literatuře a vůbec kvantifikovaně porovnává s jinými \\ \textbf{OM:} platí stále? \\ \textbf{OM:} ano!}
%\todo{\textbf{PR:} OM zavede zkratku GLP}
%
The algorithm of \cite{AtlasClark2012:Generalized.linear.prediction} represents one of the options on how to cope with the clipping consistency,
but we emphasize it is purely heuristic;
in our article, we will make an effort to formulate the constraints on the signal $\x$
in a~rigorous way and make it a~part of the modeling task.
%systematically, mathematically and justifiably.
%\todo[inline]{\textbf{OM:} Tohle bych možná taky zkusil přepsat. Sice se chceme vymezit, ale tohle na mě působí hrozně přísně. [OM]}

Other important AR-based methods for audio inpainting of compact gaps
are those based on extrapolating the right- and left-hand sides of each gap separately;
the interpolation task is then solved by crossfading the two extrapolated sequences
\cite{Kauppinen2002:reconstruction.method.long.portions.audio,Kauppinen2002:Audio.signal.extrapolation,Esquef2003:Interpolation.Long.Gaps.Warped.Burgs,Roth2003:Frequency.Warped.Burg}.
Similarly,
Etter \cite{Etter1996:Interpolation_AR} proposes to use separate AR models for the two contexts of the gap,
but suggests that the two sets of AR parameters be optimized simultaneously.
%\cite{Etter1996:Interpolation_AR}
%that extrapolates the right- and left-hand sides of each gap separately and makes the final interpolation by crossfading the two.
%\todo{\textbf{OM:} Víme, že to není úplně pravda.\\
%To je ve skutečnosti někdo jiný (upravit) a Ettera zpřesnit hned následně. Vzápětí zdůraznit, že oboje je stejně bez regularizace. [OM]}
%
%
% Přeformulovat a použít:
%
% The extrapolation methods \cite{Kauppinen2002:reconstruction.method.long.portions.audio,Kauppinen2002:Audio.signal.extrapolation,Esquef2003:Interpolation.Long.Gaps.Warped.Burgs,Roth2003:Frequency.Warped.Burg} are non-iterative and utilize a~twofold extrapolation.
%
% The patented method of Etter \cite{Etter1996:Interpolation_AR} considers the just mentioned approach suboptimal and proposes to aggregate the two extrapolation directions in a~single optimization criterion.
%
However,
neither of these contributions involves any regularization or constraints
and, most importantly,
the extrapolation-based methods require a~long reliable context of the gap, which limits the use cases significantly.

An ARMA approach to modeling time series with sparsity constraints on both the AR and MA coefficients has been presented in~\cite{VeselyToner2006sparseARMA}.
%
%A~few years later,
Later,
%\todo{\textbf{PR:} tady podle mě není jasné, po čem později}
the authors of \cite{Giacobello2009:Joint.Estimatio.Short.Long.Predictors,Giacobello2012:Sparse.Linear.Prediction}
presented their approach to audio inpainting under the name High-order sparse linear prediction (HOSpLP).
As in the follow-up papers \cite{Dufera2018:HOSpLP,Dufera2019:HOSpLP},
the goal is to recover missing samples of speech by using an AR model with sparsity regularization on the AR coefficients. 
%cílem je kombinovat short-term and long-term linear predictor
%\todo[inline,disable]{\textbf{PR:} no tady bychom se měli vůči nim nějak vyhradit, že to pořád ještě není ono.
%My jednak rozšiřujeme o regularizaci na signál, a taky ty penalizace máme obecné (teoreticky i nekonvexní). [OM] \\
%\textbf{OM:} napsal jsem [PR]}%

Importantly, none of the methods mentioned above 
offers a~flexible interconnection between the AR model and the time-domain requirements on the signal.
Furthermore,
none of the methods is suitable for audio dequantization,
where no reliable parts are available to estimate the AR model
and the time-domain constraints are essential.

A Bayesian approach to this problem has been pursued by Troughton and Godsill,
using the AR model either on the that complements a~sinusoidal deterministic component of the signal \cite{Troughton1999:Dequantization.Sinusoidal.AR},
or as a~self-contained model \cite{TroughtonGodsill1999:MCMC.restoration.quantised.time.series}.
The work is generalized in \cite{TroughtonGodsill2001:MCMC.restoration.nonlinearly.distorted.AR.signals} to other non-linear distortions.

The area of signal reconstruction has not been ignored by the advances in deep learning.
In particular, declipping and inpainting have been treated by different deep models
\cite{Tanaka2022:APPLADE,Kwon2024:Speech.declipping.transformer,Sechaud2024:Equivariance.learning.declipping,SventoRajmicMokry2024:Plug.and.play.diffusion}.
In the context of autoregression,
three approaches should be mentioned:
First,
the popular WaveNet \cite{Oord16:WaveNet} has a~conceptual connection with AR modeling,
since the generation of signal samples is conditional, given the preceding samples.
%který je koncepčně autoregresní, neboť generování časových vzorků je podmíněno vzorky předchozími.
%Ale neumí diskutovaná omezení.
%
Second,
the AR-Net \cite{Triebe2019:ARNet} represents an equivalent description of the AR model,
but from the perspective of neural networks.
In~particular, the AR coefficients are treated as network parameters and optimized using the back-propagation algorithm.
%, kde model je totožný, ale koeficienty se nehledají zaběhlými algoritmy, alébrž backpropagem.
%\todo{\textbf{PR:} Mimochodem, dá se vůbec mluvit o trénování? Nebo je to prostě jen způsob jak odhadnout koeficienty?}
This approach also allows regularizing the coefficients by simply adding terms to the loss function.
As an example, the authors of \cite{Triebe2019:ARNet} mention sparsity, creating a connection with the HOSpLP framework \cite{Giacobello2009:Joint.Estimatio.Short.Long.Predictors}.
Finally,
Mezza et al.\ propose a hybrid technique which employs an AR model in parallel with a neural network modeling the residual \cite{Mezza2024:Hybrid.packet.loss.concealment}.
However,
none of the approaches offers to \emph{constrain the time-domain signal} and direct applications to audio signal reconstruction is missing.

\subsection{Our contribution}
%\todo{přeangličtit [PR]}
%\begin{itemize}
%		\item inspirováni úspěchem GLP a HOSpLP navrhujeme neheuristické rozšíření úlohy tak, aby šlo klást podmínky na nejen na AR koeficienty, ale i na signál či oboje současně
%		\item opíráme se o optimalizační formulaci a její řešení pomocí proximálních algoritmů, což je (např. oproti GLP) lepší theoretical fundament
%%		\item bude to výpočetně náročné, ale bude to super
%\end{itemize}

Inspired by the success of the heuristic GLP approach that can apply simple constraints in the time domain,
and HOSpLP that regularizes the AR coefficients estimation with their sparsity,
we propose a~general framework that enables regularizing/constraining the time-domain and AR-domain estimation simultaneously.
Our formulation stems from a~theoretical foundation and,
compared to the above mentioned ad-hoc solutions,
it offers an optimization framework, allowing the use of standard proximal algorithms.
%our proposal does not rely on heuristics.
%\todo{PR: taky by se mohlo připsat, že je to hodně flexibilní a zahrnuje mnoho inverzních úloh. \\
%OM+PR: PR dopíše větu.}
\edt{This way, the framework can be straightforwardly applied to a much wider range of problems than the experimental part actually presents.} %PR

The rest of the paper is organized as follows:
An overview of the AR model is introduced in Sec.\ \ref{sec:AR},
together with its regularized variant.
%and with model fitting defined as an optimization task. %PR zkracuje, myslí že to není důležitá info
%The augmented AR model is presented in Sec.\ \ref{sec:AAR}, and its fitting is defined as an optimization task.
An algorithm to solve the task is developed in Sec.\ \ref{sec:algorithm}.
Sec.\ \ref{sec:experiments} demonstrates the suitability of the proposed method for the problems of audio declipping and dequantization;
the performance of the discussed approaches is compared with that of other optimization-based methods.
%Analysis of the
Effects of different parameters of the proposed model and of the algorithm are also presented.
%Finally, Sec.\ \ref{sec:conclusion} concludes the paper.  %PR zakomentoval
%Appendix~\ref{sec:acceleration} is devoted to considerations on acceleration of the numerical solver.
%Appendix~\ref{sec:Implement.details} provides implementation details.
%\todo{PR: A co ostatní přívěšky? :) \\ OM: Chceme je tu vypisovat? Možná bych řekl jenom že přívěšky obsahují nějaké bonusové obrázky, experimenty a technické detaily a hotovo.\\ PR: Proč ne :) \\ OM+PR: "Appendices A...F? provide bonus materials". PR napíše.}
Appendices \ref{sec:acceleration} to \ref{sec:Implement.details} present supplementary material.

\section{Augmenting the autoregressive model}
\label{sec:AR}

An autoregressive process of order $p$ is a~discrete-time stochastic process
$\{X_{n} \mid n = 0, \pm 1, \pm 2, \dots\}$ defined by the equation
\begin{equation}
	X_{n} + a_{2} X_{n-1} + \dots + a_{p+1} X_{n-p}  = E_{n},\quad\forall n,
	\label{eq:AR.stochastic}
\end{equation}
where $\{E_{n}\}$ is a zero-mean white noise process with variance $\sigma^2 > 0$ \cite[Def.\,3.1.2]{BrockwellDavis2006:Time.series}.
In the source--filter signal model, \eqref{eq:AR.stochastic}
%\todo{\textbf{PR:} podívat se, jak se odkazuje na rovnice v časopisu, kam budeme posílat (TASLP?)}
% PR: Kohei v TASLPu používá systematicky Eq.
% OM: pouze číslo vyjma na začátku věty, kde je celé slovo (https://journals.ieeeauthorcenter.ieee.org/wp-content/uploads/sites/7/IEEE-Editorial-Style-Manual-for-Authors.pdf, str. 24)
represents the process of passing the excitation noise $\{E_{n}\}$ through an all-pole filter with coefficients
$a_{1}, a_{2}, \dots, a_{p+1}$, where $a_{1} = 1$ is assumed without loss of generality \cite[Ch.\,4]{AudioSignalProcessingAndCoding}.

In terms of
%discrete-time signal modeling,
real-world signal modeling,
\eqref{eq:AR.stochastic} can be reformulated for a~particular observation $\x\in\RR^N$ as
\begin{equation}
	\sum_{i=1}^{p+1} a_{i} x_{n+1-i} = e_{n}, \quad a_{1} = 1,\ n = 1,\dots, N+p.
	\label{eq:error.sum}
\end{equation}
%\begin{equation}
%	% x_{n} = -a_{2} x_{n-1} - \dots - a_{p+1} x_{n-p} + e_n,
%	x_{n} + a_{2} x_{n-1} + \dots + a_{p+1} x_{n-p} = e_n,
%	\quad n = 1,\dots,N+p.
%	\label{eq:AR.signal}
%\end{equation}
The vector $\x$ is zero-padded to its new length $N+p$ such that \eqref{eq:error.sum} is correctly defined.
The definition in \eqref{eq:error.sum}
is in close relationship with linear convolution and in accordance with the convention of the \texttt{lpc} function in Matlab.
From the perspective of model fitting,
the vector $\e\in\RR^{N+p}$ is called the \emph{residual error}
% or \emph{prediction error}
\cite[Sec.\,8.2.2]{Zolzer2011:DAFX}.
%It can be expressed as
%\begin{equation}
%	e_{n} = \sum_{i=1}^{p+1} a_{i} x_{n+1-i}, \quad n = 1,\dots, N+p.
%	\label{eq:error.sum}
%\end{equation}
Given the observed signal $\x$ and the order $p$, the AR model parameters are estimated by
\begin{equation}
	\argmin_{\a} \frac{1}{2}\norm{\e(\a,\x)}^2.
	% \quad\text{subject to}\quad a_{1} = 1,
	\label{eq:AR}
\end{equation}
In \eqref{eq:AR}, we denote $ \a = \transp{[1, a_{2}, a_{3}, \dots, a_{p+1}]}$\!,
%(i.e., we inherently minimize subject to $a_{1} = 1$),
%\todo[color=orange!25!white]{\textbf{PR:} závorka se dá vyhodit, je to detail}
and $\e = \transp{[e_{1}, e_{2}, \dots, e_{N+p}]}$ is defined in \eqref{eq:error.sum} as a function of both the signal $\x$ and the coefficients $\a$.
This can effectively be solved, for example, by using the autocorrelation method and the Levinson--Durbin algorithm \cite{Durbin1960,Levinson1946}.
\edt{
Besides,
it will be later useful to rewrite the error $\e(\a,\x)$ in matrix forms as
\begin{equation}
	\e = \X\a = \A\x,
	\label{eq:error}
\end{equation}
where $\X\in\RR^{(N+p)\times (p+1)}$ and $\A\in\RR^{(N+p)\times N}$ are Toeplitz-structured,
\begin{equation}
	\label{eq:matrices}
	\A =
	\resizebox{0.75\width}{!}{$
		\begin{bmatrix}
			1 & & & \mathbf{0} \\
			a_{2} & 1 & & \\
			\vdots & a_{2} & \ddots & \\
			a_{p+1} & \vdots & & 1 \\
			& a_{p+1} & & a_{2} \\
			& & \ddots & \vdots \\
			\mathbf{0} & & & a_{p+1}
		\end{bmatrix}
		$},
	\quad
	\X =
	\resizebox{0.75\width}{!}{$
		\begin{bmatrix}
			x_{1} & & & \mathbf{0} \\
			x_{2} & x_{1} & & \\
			\vdots & x_{2} & \ddots & \\
			x_{N} & \vdots & & x_{1} \\
			& x_{N} & & x_{2} \\
			& & \ddots & \vdots \\
			\mathbf{0} & & & x_{N}
		\end{bmatrix}%
		$}.%
\end{equation}%
}%

We define the regularized autoregressive model using an optimization formulation
as follows:
\begin{equation}
	\hspace{-0.2cm}
	\argmin_{\a,\x} \left\{ Q(\a,\x) = \tfrac{1}{2}\norm{\e(\a,\x)}^2 + \lcoef \fcoef(\a) + \lsig \fsig(\x) \right\}\!,
	\hspace{-0.1cm}
	\label{eq:AAR}
\end{equation}
where we assume $\lcoef, \lsig \geq 0$, and $\fcoef, \fsig$ are (preferably) convex lower semicontinuous functions,
where the subscripts indicate the connection to the coefficients or the signal.
The idea behind \eqref{eq:AAR} is that the AR nature of the signal is preserved by minimizing the residual error
%defined above in \eqref{eq:error.sum},
as in \eqref{eq:AR}
but, at the same time,
possible priors on the AR coefficients $\a$ and the signal $\x$ are incorporated via the functions $\fcoef$ and $\fsig$, respectively.

The audio inpainting task,
\edt{as originally presented in \cite{javevr86},}
corresponds to problem \eqref{eq:AAR} with $\fcoef = 0$
and $\fsig$ being the indicator function of the set $\setinp$ of feasible signals;
see also~\eqref{eq:Inpainting.problem}.
In such a situation, % is a special case of the AAR,
the hard constraints $\Mr\x = \Mr\y$ can be directly incorporated in the noise-minimization term
\edt{%
using a complementary matrix $\Mm$ selecting the missing samples:
Decomposing
$\e(\a,\x) = \A\x = \A\transp{\Mr}\Mr\x + \A\transp{\Mm}\Mm\x$,
we arrive at
\begin{equation}
	\hspace{-0.2cm}
	\argmin_{\a,\x} \left\{ Q(\a,\x) = \tfrac{1}{2}\norm{\A\transp{\Mr}\Mr\y + \A\transp{\Mm}\Mm\x}^2 \right\}\!,
	\hspace{-0.1cm}
	\label{eq:AAR.inpainting}
\end{equation}
where we substituted $\Mr\y$ for $\Mr\x$, expressing
the hard constraint.
Given $\a$,
this task represents a least-squares problem with $\Mm\x$ being the variables, which can be solved more efficiently compared with working directly with \eqref{eq:AAR}.
}

A~similar model is the HOSpLP framework for audio inpainting \cite{Dufera2018:HOSpLP}, where the coefficient estimation is regularized solely by $\fcoef(\a) = \norm{\a}_1$, or even by a non-convex sparsity penalty \cite{Dufera2019:HOSpLP}.
In contrast to~\eqref{eq:AAR}, HOSpLP further allows utilizing a different norm for the residual $\e$, which is assumed to be more suitable, e.g., to model voiced speech
(in such a~case, the source in the source--filter model is a pulse train rather than white noise).

%\todo[inline]{\textbf{OM:} Postřeh z \texttt{oracle\_test.m}: Bez regularizace AR koeficientů se může stát, že ač se celková účelová funkce a SNR mohou zlepšovat, AR koeficienty jako takové mohou neúměrně růst (chyba v řádu $10^4$ oproti AR koeficientům čistého signálu). Pomocí parametru $\lcoef>0$ lze toto numericky nežádoucí chování ovlivnit. \\
%\textbf{PR:} lze zmínit už zde (bez odkazu na kód) jakožto další důvod pro regularizaci}

Another reason for the AR model regularization may be numerical stability.
We observed that without a~regularization of the AR coefficients, the overall objective function and SNR may improve in the course of iterations,
but the AR coefficients themselves may grow disproportionately
(for example,
a~difference
in the order of $10^4$ compared with the ground-truth signal AR coefficients can appear).
\edt{%
An extreme case we observed was a~failure of the AR model estimation for rare combinations of signals and high model orders.
In these cases, the estimation failed due to inversion of a~badly conditioned matrix.}
Using the parameter $\lcoef>0$, this numerically undesirable behavior can be controlled.
%\todo{PR: Nebylo by lepší místo indexů C,S značit prostě a,x? \\ OM: není problém, stačí změnit příkazy $\lcoef$, $\lsig$, $\gcoef$ a $\gsig$. \\ *: U lambdy a gammy by to bylo hezčí, ale muselo by se změnit i $f_C$ a $f_S$, kde by to pak vypadalo divně jako $f_a(a)$ apod.}

\section{Algorithmic approach}
\label{sec:algorithm}

Suppose the functions $\fcoef, \fsig$ are convex;
it follows that the problem \eqref{eq:AAR} poses a~biconvex optimization,
meaning that although \eqref{eq:AAR} as a~whole is non-convex,
minimizations of $Q(\a,\x)$ over $\a$ and over $\x$ separately represent convex problems.
%, i.e. a minimization of a convex function over a convex set.
%However, the problem as a whole is non-convex.
Such cases appear frequently in the literature
\cite{AharonEladBruckstein2006:KSVD,Zaviska2023:Multiple.Hankel.matrix.inpainting},
and the problem is typically treated via the Alternate Convex Search (ACS)\cite[Sec.\,4.2.1]{Gorski2007:Biconvex.sets.and.optimization}.
%\todo{\textbf{OM:} přesunuto z nižších míst}

%Note that the same update scheme
%can also be obtained from the Alternating Direction Method of Multipliers, when applied to a~biconvex problem \eqref{eq:AAR}, see \cite[Sec.\,9.2]{Boyd2011ADMM}.

%\subsection{Analysis of the problem and the ACS algorithm}
\subsection{Alternate Convex Search Algorithm}
\label{ssec:acs}
Statements regarding the convergence of the ACS are generally not possible, see also \cite[App.\,B]{javevr86}.
Thanks to biconvexity, however, at least the convergence of the objective values can be guaranteed, i.e., of the sequence $\{ Q(\a^{(i)}\!,\x^{(i)}) \mid i = 1,2,\dots\}$
\edt{\cite[Thm.\,4.5]{Gorski2007:Biconvex.sets.and.optimization}}.
%This is due to the convexity property, which results in the sequence being decreasing and bounded from below.
However, the limit of the sequence may not be the global minimum of $Q(\a,\x)$.
%and it also does not suggest any result for the sequence $\{ ( \a^{(i)}, \x^{(i)} ) \}$ generated by the ACS.
%
Some practical convergence properties can be formulated when the sequence
$\{ ( \a^{(i)}\!, \x^{(i)} ) \}$ is contained in a~compact set,
%which could be enforced by the functions $\fcoef, \fsig$,
%\todo{\textbf{PR:} vágní}
which could be enforced, for example, by defining $\fcoef, \fsig$ as indicator functions \cite{combettes2011proximal},
and if both the individual convex sub-problems
%\eqref{eq:ACS2}
have unique solutions.
Under these assumptions, the ACS sequence 
%generated by the ACS
has at least one accumulation point and all the accumulation points are stationary points of $Q(\a,\x)$
\cite[Thm.\,4.9, Corr.\,4.10]{Gorski2007:Biconvex.sets.and.optimization},
\edt{\cite[Sec.\,4.2]{JainKar2017:Non-convex.optimization}.}
\edt{Such a~behavior is reported also in \cite[App.\,B]{javevr86}.} %PR
\edt{Reference \cite{javevr86}
%also discusses the convergence speed, and
presents a relation to the expectation--maximization scheme (EM) which, however, is not accompanied with encouraging convergence properties in a general setup
\cite[Ch.\,5]{JainKar2017:Non-convex.optimization}.} %PR

%The main idea of the algorithm stems from the biconvexity property---it consists of convex optimization sub-problems, which are obtained by fixing one of the variables incorporated in the whole problem, and minimizing with respect to the other variable.
%\todo{\textbf{PR:} moc podrobné}
%
%In the case of \eqref{eq:AAR},
%the ACS algorithm can be summarized as iterating the following two steps,
%after an initial $\x^{(0)}\in\RR^N$ and the AR model order $p$ are set:
%%
%\begin{subequations}
%	\label{eq:ACS1}
%	\begin{align}
%		\a^{(i)}
%		&= \argmin_{\a\in\RR^{p+1}\!,\,a_1=1} Q(\a,\x^{(i-1)}) %,
%		\label{eq:ACS1:a}\\
%		\x^{(i)}
%		&= \argmin_{\x\in\RR^N}\ Q(\a^{(i)}\!,\x) %,
%		\label{eq:ACS1:x}
%	\end{align}
%\end{subequations}
%%
%for $i = 1,2,\dots$, until a~termination criterion is met.
%%
%\todo{\textbf{PR:} tento a předchozí odstavec lze ještě více zkrátit a sloučit \\ OM: nevymyslel jsem jak, ale rovnice \eqref{eq:ACS1} by šly ří}
%In both the sub-problems \eqref{eq:ACS1:a} and \eqref{eq:ACS1:x},
%one of the variables
%%$\a$ and $\x$
%plays the role of as a constant;
%therefore,
%%part of the objective function $Q(\a,\x)$ can be omitted and
%the two simplify to
%
%\todo{\textbf{OM:} provedeno drobné zkrácení}
In both the sub-problems of minimizing
%$Q(\a,\x)$
$Q$
over $\a$ and over $\x$,
the other variable is constant, which allows omitting the respective part of the objective function in \eqref{eq:AAR}
and simplifying the updates to
\begin{subequations}
	\label{eq:ACS2}
	\begin{align}
		\a^{(i)}
		&= \argmin_{\a} \left\{ \tfrac{1}{2}\norm{\e(\a,\x^{(i-1)})}^2 + \lcoef \fcoef(\a) \right\}\!,
		\label{eq:ACS2:a}\\
		\x^{(i)}
		&= \argmin_{\x} \left\{ \tfrac{1}{2}\norm{\e(\a^{(i)}\!,\x)}^2 + \lsig \fsig(\x) \right\}\!.
		\label{eq:ACS2:x}
	\end{align}
\end{subequations}

In the following subsections, the ways to solve
%the sub-problems
\eqref{eq:ACS2:a} and \eqref{eq:ACS2:x} are discussed.
As will be demonstrated,
%later in Sec.\ \ref{sec:experiments},
it turns out that the bottleneck
%of the method
is the computational complexity.
Therefore, the possibilities of accelerating the ACS procedure will also be discussed.
% in Appendix \ref{sec:acceleration}.
% OM: nejen v Appendixu

\subsection{Dealing with the sub-problems}
\label{sec:algorithms.for.subproblems}

In the convex setting and with
%nontrivial
nonzero
%\todo{\textbf{PR:} ? \\ OM: myslí se nenulové}
$\fcoef, \fsig$, the problems \eqref{eq:ACS2:a} and \eqref{eq:ACS2:x} can be addressed via proximal splitting methods~\cite{combettes2011proximal}.
Each of the problems consists of a~sum of two convex functions:
the quadratic AR model error and a possibly non-smooth penalty $\fcoef$ or $\fsig$.
%one of which
%\todo{PR. To je vždycky ta druhá, ano?}
%is possibly non-smooth.
The key concept of the proximal splitting is to reach the global optimum 
%divide the objective into a sum of several functions
%(if it is not in such a form due to the nature of the problem)
by repetitive minimization using gradients or the so-called proximal operators related to individual summands.

% ======================================
% NOVÁ VERZE O VOLBĚ PROXIMÁLNÍHO ALGORITMU
% =========================================

The proximal operator of a~function $f$ is another function acting on the same space, defined as
%\footnote{%
%If the function $f$ is convex lower semicontinuous,
%the minimization has a~unique solution, thus
%the corresponding proximal operator is well defined.
%}
%\todo{PR: footnote mi přijde zbytečná v našem článku}
$\prox_{f}(\uu) = \argmin_{\vv} \{\frac{1}{2}\norm{\vv - \uu}^2 + f(\vv)\}$.
Among the typical examples is the projection onto a convex set $C$,
which is the proximal operator of the indicator function of the set, $\iota_C$.
Another example is the soft thresholding with the threshold $\gamma$ as the proximal operator of the scaled $\ell_1$ norm $\gamma\norm{\cdot}_1$.
For more examples and background, see e.g.\ \cite{combettes2011proximal}, \cite[Ch.\,6]{Beck2017:First.Order.Methods}.

Each of the problems \eqref{eq:ACS2:a} and \eqref{eq:ACS2:x}
can be addressed via
the Douglas--Rachford algorithm (DRA) \cite[Sec.\,4]{combettes2011proximal},
which requires access to the proximal operators of the two summands involved.
In the case of \eqref{eq:ACS2}, this applies to the quadratic function and either $\lcoef\fcoef$ or $\lsig\fsig$.

We further assume that $\prox_{\lcoef \fcoef}(\a)$ and
$\prox_{\lsig \fsig}(\x)$ are available in an explicit form,
which will be the case later in Section \ref{sec:experiments} (see also Appendix \ref{sec:Implement.details}).
To deal with the quadratic model error term,
\edt{recall the matrix form presented in \eqref{eq:error}.}
Then, the proximal operators of $\frac{1}{2}\norm{\e(\a,\x)}^2$ with respect to $\a$ and $\x$
can simply be expressed as \cite[Table\,I]{combettes2011proximal}
\begin{subequations}
	\label{eq:prox.error}
	\begin{align}
		\prox_{\gcoef\norm{\e(\cdot, \x)}^2/2}(\a) &= (\Id + \gcoef\transp{\X}\!\X)^{-1}\a,
		\label{eq:prox.error:a}
		\\
		\prox_{\gsig\norm{\e(\a, \cdot)}^2/2}(\x) &= (\Id + \gsig\transp{\A}\!\A)^{-1}\x.
		\label{eq:prox.error:x}
	\end{align}
\end{subequations}
Here $\gcoef$ and $\gsig$ are parameters of the DRA for the respective problems.
% and we use the expression for the error from \eqref{eq:error}.
\edt{%
The underlying structure of the matrices allows an efficient numerical implementation without employing the inverse explicitly---it is based on
% embedding the matrices $\X$ and $\A$ in the problems
% \eqref{eq:ACS2:a} and \eqref{eq:ACS2:x},
appending rows to the matrices $\X$ and $\A$ in 
\eqref{eq:prox.error:a} and \eqref{eq:prox.error:x},
respectively,
yielding a~circulant structure, allowing obtaining the inverses
%into larger matrices, for which the multiplication with the aforementioned inverses can be computed
efficiently using the FFT.
Importantly,
applying these extended proximal operators in solving the problems \eqref{eq:ACS2:a} and \eqref{eq:ACS2:x} guarantees reaching the same minimum as with employing \eqref{eq:prox.error:a} and \eqref{eq:prox.error:x} directly}
% the extended problem is guaranteed to share its minimum with the original problem
\cite{Bayram2015:Structured}.
%However,
%the problem being solved is necessarily of a~larger dimension, which may result in different convergence properties.

Note that further acceleration is achievable by heuristic schemes---an example is the extrapolation of the updates given by \eqref{eq:ACS2}.
These alternatives, as well as a larger evaluation of the properties of the FFT-accelerated DRA in the context of audio declipping, can be found in Appendix \ref{sec:acceleration}.

% ===============================================
% KONEC NOVÉ VERZE O VOLBĚ PROXIMÁLNÍHO ALGORITMU
% ===============================================

\edt{%
The whole ACS is summarized in Algorithm \ref{algorithm}, featuring the Janssen audio inpainting \cite{javevr86} as a particular case.
}

\begin{algorithm}
	\DontPrintSemicolon
	\rightskip=-5mm % OM: trik pro využití celé šířky pole
	\caption{%
		\edt{%
			ACS solving \eqref{eq:AAR}
		}
	}
	\label{algorithm}
	\edt{%
		initialize $\x^{(0)}$ \\
		\For{$i=1,2,\dots$}{
			\fbox{coef.\ update \eqref{eq:ACS2:a}} \\
			define $\X$ given $\x^{(i-1)}$
			\tcp*{\eqref{eq:matrices}}
			\eIf{$\lcoef=0$}{
				$\a^{(i)} = [1; \X(\texttt{:,2:end})^+\X\texttt{(:,1)}]$
				\label{ln:lpc}
				\tcp*{\texttt{lpc($\x^{(i)}$)}}
			}
			{
				define $\prox_{\gcoef \lcoef \fcoef}(\a)$
				\label{ln:prox.coef}
				\tcp*{Appendix \ref{sec:Implement.details}}
				$\prox_{\gcoef\norm{\e(\cdot, \x)}^2/2}(\a) = (\Id + \gcoef\transp{\X}\!\X)^{-1}\a$
				\tcp*{\eqref{eq:prox.error:a}}
				$\a^{(i)} = \textup{DRA}(\prox_{\gcoef\norm{\e(\cdot, \x)}^2/2}, \prox_{\gcoef \lcoef \fcoef}, \textit{params})$
			}
			extrapolate the estimate $\a^{(i)}$
			\label{ln:acceleration.a}
			\tcp*{Appendix \ref{sec:acceleration}}
			\fbox{signal update \eqref{eq:ACS2:x}} \\
			define $\A$ given $\a^{(i)}$
			\tcp*{\eqref{eq:matrices}}
			\eIf{inpainting, consistent}{
				set $\Mr\x^{(i)} = \Mr\y$ \\
				set $\Mm\x^{(i)} = -(\A\transp{\Mm})^+\A\transp{\Mr}\Mr\y$
				\label{ln:inpainting}
				\tcp*{\eqref{eq:AAR.inpainting}}
			}
			{
				define $\prox_{\gsig \lsig \fsig}$
				\label{ln:prox.sig}
				\tcp*{Appendix \ref{sec:Implement.details}} 
				$\prox_{\gsig\norm{\e(\a, \cdot)}^2/2}(\x) = (\Id + \gsig\transp{\A}\!\A)^{-1}\x$
				\tcp*{\eqref{eq:prox.error:x}}
				$\x^{(i)} = \textup{DRA}(\prox_{\gsig\norm{\e(\a, \cdot)}^2/2}, \prox_{\gsig \lsig \fsig}, \textit{params})$
			}
			extrapolate the estimate $\x^{(i)}$
			\label{ln:acceleration.x}
			\tcp*{Appendix \ref{sec:acceleration}}
			perform line search
			\label{ln:acceleration.ls}
			\tcp*{Appendix \ref{sec:acceleration}}
		}
	}
\end{algorithm}

\edt{%
Line \ref{ln:lpc} of Algorithm \ref{algorithm} uses Matlab notation for the selection of matrix rows and columns.
The formula itself can be derived from \eqref{eq:AR} by substituting $\e=\X\a$,
see \eqref{eq:error} and \eqref{eq:matrices}, fixing $\a_1=1$ and solving the resulting least-squares problem using the pseudoinverse, denoted as $(\cdot)^+$\!.
Similarly, line \ref{ln:inpainting} presents the explicit solution to the least-squares problem \eqref{eq:AAR.inpainting}.
}

\edt{%
Note that in practice, the proximal operators on lines \ref{ln:prox.coef} and \ref{ln:prox.sig} do not depend on the current value of $\x^{(i-1)}$ and $\a^{(i)}$\!, respectively.
Also note that the acceleration steps on lines \ref{ln:acceleration.a}, \ref{ln:acceleration.x} and \ref{ln:acceleration.ls} are optional and heuristic.
}

\section{Experiments and results}
\label{sec:experiments}

This section mostly concerns the audio declipping task;
the application to audio dequantization is discussed in Sec.\,\ref{ssec:dequantization}.

The AR-based declipping task
can be posed as an optimization problem in the form of \eqref{eq:AAR}: %, i.e.,
\begin{equation}
	\argmin_{\a,\x} \left\{ \tfrac{1}{2}\norm{\e(\a,\x)}^2 + \lcoef\norm{\a}_1 + \lsig d_{\setdec}(\x)^2 / 2\right\},
	\label{eq:AAR.declipping}
\end{equation}
where $d_{\setdec}(\x) = \norm{\x-\proj_{\setdec}(\x)}_2$ denotes the distance of $\x$ from the set of feasible signals $\setdec$.
Letting \mbox{$\lsig=\infty$}
%\todo[inline]{\textbf{PR:} Jak to děláš v praxi? Nebo je to jen formální zápis? \\ \textbf{OM:} formalita, v praxi se liší prox pro $\lsig=\infty$ (projekce) a pro $\lsig<\infty$\\
%*: Dopsat appendix o použití dvou variant; možná že $\infty$ (projekce) je dokonce limitním případem [OM]}%
allows the problem \eqref{eq:AAR.declipping} to cover the \emph{consistent} declipping solution, 
%where we enforce
yielding %PR
the hard constraint $\x\in\setdec$ as in \eqref{eq:declipping.problem.generally}
(see also Appendix \ref{sec:Implement.details}).
In the case of $0 < \lsig < \infty$, the solution is allowed to be \emph{inconsistent}
(see also Fig.\ \ref{fig:waveforms.consistency}).
%However, strategies exist to leverage the samples that are not clipped \cite{ZaviskaRajmicMokry2022:Declipping.crossfading}.
%\todo{PR: O crossfadingu se mluví v exp. sekci, zde bych klidně vyhodil, je to rušivé a myslím nepodstatné}

Furthermore,
we compare the solution to our problem \eqref{eq:AAR.declipping} with two
reduced %PR
%simplified
approaches.
The first one is the inpainting strategy from the original work of Janssen et al.\ \cite{javevr86}, treating the clipped samples as completely missing with no clipping-related constraints.
%\todo{PR: Ukončit větu tímto?: corresponding to lambdaS=0 \\ OM: To není celá pravda, protože podmínku na signál máme, ale jenom na R pozicích. Radši bych nezabrušoval.\\ PR: Nezabrušujme tudíž :)}
The second one is the generalized linear prediction (GLP),
described in Sec.\ \ref{sec:AR.methods.for.signal.reconstruction},
which heuristically modifies Janssen's original approach to declipping \cite{AtlasClark2012:Generalized.linear.prediction}.

Following on the HOSpLP framework \cite{Dufera2018:HOSpLP}, we allow the regularization of the AR model coefficients via the $\ell_1$ norm by setting $\lcoef>0$ in \eqref{eq:AAR.declipping}.
The same regularization is applied
in the inpainting and GLP cases, where the AR model update \eqref{eq:ACS2:a} does not differ from the declipping case.

To deal with long (and non-stationary) signals, the proposed regularized AR model is used and estimated frame-wise.
%After optimization, the resulting audio signal is formed using the overlap-add strategy.
The same as in the seminal inpainting paper~\cite{Adler2012:Audio.inpainting},
we use the rectangular window to divide the input signal into overlapping
segments of length $w$;
after optimization, the final overlap-add synthesis is done with the sine window to enable smooth transition between adjacent frames.

%\todo[inline]{\textbf{Poznámka k volbě okna:} Janssen dle Adlera používá obdélníkové okno + OLA se sinusovým \cite[Tab.\,III]{Adler2012:Audio.inpainting}.\\
%\textbf{PR:} OK. Mělo by se zdůraznit i v textu, aby nebyla mýlka, že děláme klasické okénkování.
%Úplně ideálně odůvodnit že to jeho OLA je správná syntéza k obdélníkové analýze :)}

Throughout this section, the quality of the declipped audio signal is assessed using the signal-to-distortion ratio (SDR).
For the original (undegraded) signal $\y$ and a reconstruction $\hat{\x}$, SDR in decibels is computed as
\begin{equation}
	\textup{SDR}(\y, \hat{\x}) = 10\log_{10}\frac{\norm{\y}^2}{\norm{\y-\hat{\x}}^2}\!.
	\label{eq:sdr}
\end{equation}
As in the declipping survey \cite{ZaviskaRajmicOzerovRencker2021:Declipping.Survey},
the perceived quality of the signal is additionally evaluated using the PEMO-Q \cite{Huber:2006a}
and PEAQ metrics \cite{Thiede:2000a, Kabal2002:PEAQ}.
Both of them predict the subjective difference of the signals $\y$ and $\hat{\x}$ in terms of the objective difference grade (ODG),
which ranges from $-4$ (very annoying impairment) to 0 (imperceptible difference).

%\todo[color=orange!25!white]{\textbf{OM:} Přepočítat s verzí PEMO-Q 1.4.1? \\ \textbf{OM:} Šlo by s pomocí skriptu \texttt{survey\_test\_add\_CR.m}, ale nevím co pro referenční metody (vzali jsme ze Survey přímo výsledná ODG). \\ \textbf{OM:} všechny zvuky jdou stáhnout z \url{https://github.com/rajmic/declipping2020/tree/master/sounds} a znova ohodnotit\\
%\textbf{PR:} je to jen otázkou časové náročnosti; já bych se tím asi nezdržoval\\
%*: Teď nechat; na github pak připsat poznámku o verzi 1.4.0}

The test signals are also taken from the survey \cite{ZaviskaRajmicOzerovRencker2021:Declipping.Survey}, i.e.,
%musical recordings sampled at 44.1\,kHz and cropped to length of around 7 seconds are used, based on the EBU SQAM database \cite{EBUSQAM}.
%\todo[color=orange!25!white]{\textbf{OM:} zobrazuje se škaredě v seznamu literatury}
%In the comparison with the survey results (subsection \ref{ssec:comparison.survey}), the corresponding set of 10 signals is used.
%%In some of the preliminary experiments, only a single violin recording is used.
%%or a subset of 5 signals from the test set are used.
%On the other hand,
%the preliminary experiments in \ref{ssec:iteration.tradeoff}--\ref{ssec:regularization} are limited to only a~subset of the test signals.
musical recordings sampled at 44.1\,kHz and cropped to a~length of around 7~seconds are used, originally based on the EBU SQAM database \cite{EBUSQAM}.
%\todo[color=orange!25!white]{\textbf{OM:} zobrazuje se škaredě v seznamu literatury\\ PR: Pro první verzi bych nechal být.}
%In the comparison with the survey results (subsection \ref{ssec:comparison.survey}), the corresponding set of 10 signals is used.
%%In some of the preliminary experiments, only a single violin recording is used.
%%or a subset of 5 signals from the test set are used.
%On the other hand,
%the preliminary experiments in \ref{ssec:iteration.tradeoff}--\ref{ssec:regularization} are limited to only a~subset of the test signals.

In the following subsections \ref{ssec:iteration.tradeoff} to \ref{ssec:regularization},
preliminary experiments focused on various aspects of the numerical optimization are presented;
these are performed on limited subsets of the above described dataset.
The main experiment \edt{on declipping of musical signals} comes in Sec.~\ref{ssec:comparison.survey},
where the methods considered are evaluated against the results of the declipping survey;
%\cite{ZaviskaRajmicOzerovRencker2021:Declipping.Survey};
there, the full set of signals is used.
\edt{Extension to declipping of speech signals is provided in Sec.~\ref{ssec:speech}, followed by illustration on the dequantization problem (Sec.~\ref{ssec:dequantization}) and a~discussion of computational complexity (Sec.~\ref{ssec:complexity}).}
%\todo{PR: psal bych jenom Sec.}

%\todo[inline,disable]{\textbf{OM:} Někde musí zaznít, že klipujeme buď na dané hladině vzhledem k maximální výchylce (např.\ 0.2), nebo podle input SDR.
%Protože pak se to používá v některých grafech.\\
%\textbf{PR:} Jo, mluví se o tom na začátku \ref{ssec:comparison.survey}. \\
%\textbf{OM:} A stačí to tak?\\ *: ano}

\subsection{Trade-off between Outer and Inner Iteration Counts}
\label{ssec:iteration.tradeoff}

\todo[inline,disable]{\textbf{PR:} Vlastně bych tady připsal předradličku ve smyslu, že teď bude pár nezáživných experimentů, a koho zjímá kvalitativní srovnání s konkurenty, ať skočí rovnou na sekci XYZ :) \\ \textbf{OM:} Je to celkem zřejmé z předchozí věty. Chceme to víc zdůraznit?\\
*: Nechme jen tam, ale PR přepíše poslední 2 věty.\\
PR: Přepsáno.}%
Since the algorithmic approach proposed in Section \ref{sec:algorithms.for.subproblems} involves nested iterations,
a natural question arises on the trade-off between the number of iterations of the whole ACS (outer)
and the number of iterations of the sub-solver (inner).
To examine this trade-off,
we perform the declipping of a~single signal frame by solving problem \eqref{eq:AAR.declipping}.
For different combinations of the number of inner and outer iterations, the reconstruction quality is measured using the SDR,
together with tracking the value of the objective function of \eqref{eq:AAR.declipping}.

\begin{figure}
	\centering
	\scalebox{0.65}{\input{figures/iteration_tradeoff_sdr.tex}} \\[1.0ex]
	\scalebox{0.65}{\input{figures/iteration_tradeoff_objective.tex}}
	\caption{Trade-off between the count of inner and outer iterations.
		The problem being solved is the consistent declipping of a single signal frame,
		i.e., problem~\eqref{eq:AAR.declipping} with $\lsig=\infty$ and varying $\lcoef$.
		The signal segment length is 2048 samples, the model order is $p=512$.
		For reference, the SDR of the clipped signal is 4.3\,dB.}
	\label{fig:tradeoff}
	\vspace{1ex}
\end{figure}

\edt{%
	The main conclusion from this experiment is that a~sufficient number of both inner and outer iterations is crucial for reaching a~low objective value/high SDR.
	However, from approximately 100 outer and 1000 inner iterations upwards, the result no longer improves.
	Importantly, the consistent improvement with the number of outer iterations (i.e., along the horizontal axis) suggests that the unknown convergence properties, as discussed in Sec.\ \ref{ssec:acs}, are not very limiting, at least in our setup.
	The last conclusion is that especially in the case of AR model regularization (plots in the right column of Fig.\ \ref{fig:tradeoff} with $\lcoef=0.01$),
	%\todo{PR: slovo "appears" není nic moc :)}
	the number of inner iterations (i.e., the precision in solving sub-problems \eqref{eq:ACS2}) is critical.
}

%The main conclusion from this experiment is that for a~large number of ACS iterations, the result no longer improves.
%Especially in the case of AR model regularization (plots in the right column of Fig.\ \ref{fig:tradeoff} with $\lcoef=0.01$),
%no significant improvement is observed from 10 \edt{outer} iterations upwards.
%Interestingly, this is in correspondence with \cite{javevr86}, where even a~single (outer)
%iteration was considered sufficient.
%This also means that the unknown convergence properties, as discussed in Sec.\ \ref{ssec:acs}, seem not very limiting, at least in our setup.
%Contrarily, the number of inner iterations (i.e., the precision in solving sub-problems \eqref{eq:ACS2})
%appears crucial.
%This interpretation stems from the observation that both the SDR and the objective function in the plots in Fig.\ \ref{fig:tradeoff} vary significantly in the vertical direction.

\subsection{Progression towards ground-truth coefficients?}

\begin{figure*}[ht]
	\centering
	\scalebox{0.77}{\input{figures/oracle.tex}} % oracle_test_14.m
	\caption{Analysis of convergence of the three main competitors towards both the clean signal and the ground-truth AR coefficients.
		In this particular case, a~$2048$-sample-long realization of an AR process of the order $p=128$ was used as the ground-truth signal and clipped at $0.2$ times the peak value (corresponding to an SDR of 4.6\,dB on the clipped samples).
		For reconstruction, we set $\lcoef = 0.1$ and $\lsig=\infty$
		(i.e., declipping in its consistent variant).
		All three algorithms run for $100$ outer iterations and $1\,000$ inner iterations of the (non-accelerated) DRA.
		Note that the objective functions (the first graph) are in general different for different algorithms due to the choices of the parameters $\lcoef, \lsig$.}
	\label{fig:oracle.test}
\end{figure*}

The previous experiment suggests
the suitability of the regularized AR model for audio declipping,
since the SDR values indicate that the reconstructed audio resembles the undistorted original.
%\todo{\textbf{PR:} Z čeho to má čtenář indikovat? Z vysokého SDR ve figuře?}
This section aims at answering the subsequent question,
whether even the sequence of AR coefficients $\{\a^{(i)}\}$ generated by the ACS converges towards the AR coefficients $\a^{\textup{true}}$ of the original signal (prior to clipping).

To investigate this subject,
a~declipping experiment
according to \eqref{eq:AAR.declipping}
is performed, where,
in addition to the objective and the SDR,
we track the evolution of the (regularized) AR coefficients in the iterations of the ACS.
This allows a~comparison with the coefficients of the undistorted signal
using the norm $\norm{\a^{(i)}-\a^{\textup{true}}}$.

The results of the experiment are presented in Fig.\ \ref{fig:oracle.test}.
The conclusion is that
it is possible that the AR coefficients converge over iterations%
---usually the relative difference between iterations decreases, see the last plot in Fig.\ \ref{fig:oracle.test} showing the values of $\norm{\a^{(i)}-\a^{(i-1)}}/\norm{\a^{(i)}}$.
%However, it appears that the coefficients of the undamaged signal are not the potential limit, because the norm of the difference to this \emph{actual} solution (plotted in the third graph) usually does not decrease.
However, is appears that
\edt{%
	even if the sequence $\{\a^{(i)}\}$ converges, it does not converge to the coefficients of the undamaged signal $\a^{\textup{true}}$.
	This is
}
because the norm of the difference to this \emph{actual} solution (plotted in the third graph) usually does not decrease.

\subsection{Regularization}
\label{ssec:regularization}

% ====================================================
% ZAČÁTEK NOVÉ VERZE O VOLBĚ REGULARIZAČNÍCH PARAMETRŮ
% ====================================================

After demonstrating that the proposed algorithmic approach allows fitting the regularized AR model,
the aim is to dig deeper in the declipping application.
A common question in regularized inverse problems is the trade-off related to the regularization strength,
typically governed by a~scalar parameter.
In the case of the generic problem \eqref{eq:AAR}, or the particular declipping formulation \eqref{eq:AAR.declipping},
there are two regularization parameters $\lcoef,\lsig$, which may significantly affect the reconstruction quality.

To evaluate the effect of the regularization,
an experiment is performed
where clipped signals are reconstructed using all three strategies
(inpainting, GLP, declipping),
for different choices of $\lcoef$ and $\lsig$ in the case of the proposed declipping approach.
The results, in terms of SDR, are presented in Table~\ref{tab:lambdas.SDR}.
The observation is that the results are not very sensitive to the choice of $\lsig$,
even though the declipping strategy (i.e., $\lsig>0$) leads to a~minor improvement over inpainting or GLP.
Regarding $\lcoef$, the choice to regularize the AR coefficients (i.e., $\lcoef>0$) also improves the reconstruction quality;
however, the regularization must be chosen carefully in order not to be overly strong
(see the results for $\lcoef=10^{-1}$).

\begin{table*}
	\centering
	\caption{Mean SDR (dB) from 10 signals, each initially degraded by clipping at an input SDR of 10 dB.
		The parameters used in the reconstruction are $w=2048$, $p=512$, 10 outer iterations and 1\,000 inner iterations.
		For each algorithm (declipping, GLP, inpainting), the maximal value (prior to rounding) is highlighted.
	}
	\vspace{-1.5mm}
	\label{tab:lambdas.SDR}
	\begin{tabular}{lccccccc@{}}
		\toprule
		& $\lcoef = 0$ & $\lcoef=10^{-6}$ & $\lcoef=10^{-5}$ & $\lcoef=10^{-4}$ & $\lcoef=10^{-3}$ & $\lcoef=10^{-2}$ & $\lcoef=10^{-1}$ \\ \midrule
		declipping, $\lsig = \infty$ & $22.9$ & $23.5$ & $23.5$ & $23.7$ & $24.3$ & $23.1$ & $19.1$ \\
		declipping, $\lsig = 100$ & $22.9$ & $23.5$ & $23.5$ & $23.7$ & $\mathbf{24.3}$ & $23.1$ & $19.1$ \\
		declipping, $\lsig = 10$ & $22.9$ & $23.5$ & $23.6$ & $23.7$ & $24.3$ & $23.0$ & $18.9$ \\
		declipping, $\lsig = 1$ & $22.4$ & $23.1$ & $23.1$ & $23.3$ & $23.8$ & $22.1$ & $18.0$ \\
		GLP & $22.4$ & $22.4$ & $22.4$ & $22.5$ & $\mathbf{22.6}$ & $21.9$ & $18.0$ \\
		inpainting & $23.4$ & $23.2$ & $23.3$ & $23.5$ & $\mathbf{23.9}$ & $22.1$ & $17.7$ \\ \bottomrule
	\end{tabular}	
\end{table*}
\subsection{\edt{Declipping of musical signals}}
\label{ssec:comparison.survey}
%\todo{Možná přejmenovat (zdůraznit music, protože přibyla řeč).}

To provide a context for the reconstruction quality reachable by the regularized AR model,
our last experiment compares the results with the survey \cite{ZaviskaRajmicOzerovRencker2021:Declipping.Survey}.
The same set of 10 musical signals is used and degraded in the same way, i.e.,
hard-clipped with the clipping level chosen according to a~given input SDR
(5, 7, 10 and 15\,dB). %přidal PR

The order of the regularized AR model is $p=512$, with the frame length set to $w=8192$ samples.
The inner iterations are solved using the FFT-accelerated DRA with 1\,000 iterations,
and the ACS algorithm runs for 5~outer iterations with signal extrapolation (see Appendix \ref{ssec:extrapolation}).

\begin{figure*}
	\subfloat[
		comparison in terms of $\Delta$SDR, i.e.\ the improvement of SDR over the clipped signal
	]{
		\hspace{-0.72in} % OM: schválně jinak, aby sedělo horizontální zarovnání
		\adjustbox{scale=0.5}{% This file was created by matlab2tikz.
\definecolor{mycolor1}{rgb}{0.72998,0.69059,0.74406}
\definecolor{mycolor2}{rgb}{0.45997,0.38117,0.48811}
\definecolor{mycolor3}{rgb}{0.18995,0.07176,0.23217}
\definecolor{mycolor4}{rgb}{0.75123,0.75442,0.88469}
\definecolor{mycolor5}{rgb}{0.50246,0.50885,0.76937}
\definecolor{mycolor6}{rgb}{0.25369,0.26327,0.65406}
\definecolor{mycolor7}{rgb}{0.75897,0.81382,0.97109}
\definecolor{mycolor8}{rgb}{0.51794,0.62763,0.94219}
\definecolor{mycolor9}{rgb}{0.27691,0.44145,0.91328}
\definecolor{mycolor10}{rgb}{0.74809,0.86979,0.99899}
\definecolor{mycolor11}{rgb}{0.49618,0.73958,0.99798}
\definecolor{mycolor12}{rgb}{0.24427,0.60937,0.99697}
\definecolor{mycolor13}{rgb}{0.18995,0.07176,0.23217}
\definecolor{mycolor14}{rgb}{0.25369,0.26327,0.65406}
\definecolor{mycolor15}{rgb}{0.27691,0.44145,0.91328}
\definecolor{mycolor16}{rgb}{0.24427,0.60937,0.99697}
\definecolor{mycolor17}{rgb}{0.13278,0.77165,0.88580}
\definecolor{mycolor18}{rgb}{0.10342,0.89600,0.71500}
\definecolor{mycolor19}{rgb}{0.27597,0.97092,0.51653}
\definecolor{mycolor20}{rgb}{0.53255,0.99919,0.30581}
\definecolor{mycolor21}{rgb}{0.72596,0.96470,0.20640}
\definecolor{mycolor22}{rgb}{0.88331,0.86553,0.21719}
\definecolor{mycolor23}{rgb}{0.98000,0.73000,0.22161}
\definecolor{mycolor24}{rgb}{0.99297,0.55214,0.15417}
\definecolor{mycolor25}{rgb}{0.94084,0.35566,0.07031}
\definecolor{mycolor26}{rgb}{0.83926,0.20654,0.02305}
\definecolor{mycolor27}{rgb}{0.68602,0.09536,0.00481}
\definecolor{mycolor28}{rgb}{0.47960,0.01583,0.01055}
\begin{tikzpicture}[font=\large]

\begin{axis}[%
width=11in,
height=5in,
at={(0in,0in)},
scale only axis,
bar shift auto,
xmin=0.5,
xmax=4.5,
xtick={1,2,3,4},
xticklabels={{5},{7},{10},{15}},
xlabel style={font=\large},
xlabel={input SDR (dB)},
ymin=-1,
ymax=22,
ylabel style={font=\large},
ylabel={$\Delta$SDR (dB)},
ytick={0,5,10,15,20,25},
yminorticks=true,
minor y tick num=5,
ymajorgrids,
yminorgrids,
axis background/.style={fill=white},
legend style={
	at={(1.03,1.0)},
	anchor=north west,
	legend cell align=left,
	align=left,
	/tikz/column 1/.style={
		column sep=4pt,
}}
]

\addlegendimage{draw=black, fill=mycolor1, area legend}
\addlegendentry{inp., $\lcoef= 0$}

\addlegendimage{draw=black, fill=mycolor2, area legend}
\addlegendentry{inp., $\lcoef = 10^{-5}$}

\addlegendimage{draw=black, fill=mycolor3, area legend}
\addlegendentry{inp., $\lcoef= 10^{-3}$}

\addlegendimage{draw=black, fill=mycolor4, area legend}
\addlegendentry{GLP, $\lcoef= 0$}

\addlegendimage{draw=black, fill=mycolor5, area legend}
\addlegendentry{GLP, $\lcoef = 10^{-5}$}

\addlegendimage{draw=black, fill=mycolor6, area legend}
\addlegendentry{GLP, $\lcoef= 10^{-3}$}

\addlegendimage{draw=black, fill=mycolor7, area legend}
\addlegendentry{dec., $\lcoef= 0$, $\lsig= 10$}

\addlegendimage{draw=black, fill=mycolor8, area legend}
\addlegendentry{dec., $\lcoef = 10^{-5}$, $\lsig= 10$}

\addlegendimage{draw=black, fill=mycolor9, area legend}
\addlegendentry{dec., $\lcoef= 10^{-3}$, $\lsig= 10$}

\addlegendimage{draw=black, fill=mycolor10, area legend}
\addlegendentry{dec., $\lcoef= 0$, $\lsig= \infty$}

\addlegendimage{draw=black, fill=mycolor11, area legend}
\addlegendentry{dec., $\lcoef = 10^{-5}$, $\lsig= \infty$}

\addlegendimage{draw=black, fill=mycolor12, area legend}
\addlegendentry{dec., $\lcoef= 10^{-3}$, $\lsig= \infty$}

\addlegendimage{draw=black, fill=mycolor13, area legend}
\addlegendentry{C-OMP}

\addlegendimage{draw=black, fill=mycolor14, area legend}
\addlegendentry{A-SPADE}

\addlegendimage{draw=black, fill=mycolor15, area legend}
\addlegendentry{S-SPADE}

\addlegendimage{draw=black, fill=mycolor16, area legend}
\addlegendentry{CP}

\addlegendimage{draw=black, fill=mycolor17, area legend}
\addlegendentry{DR}

\addlegendimage{draw=black, fill=mycolor18, area legend}
\addlegendentry{SS EW}

\addlegendimage{draw=black, fill=mycolor19, area legend}
\addlegendentry{SS PEW}

\addlegendimage{draw=black, fill=mycolor20, area legend}
\addlegendentry{CSL1}

\addlegendimage{draw=black, fill=mycolor21, area legend}
\addlegendentry{PCSL1}

\addlegendimage{draw=black, fill=mycolor22, area legend}
\addlegendentry{PWCSL1}

\addlegendimage{draw=black, fill=mycolor23, area legend}
\addlegendentry{reweighted CP}

\addlegendimage{draw=black, fill=mycolor24, area legend}
\addlegendentry{reweighted DR}

\addlegendimage{draw=black, fill=mycolor25, area legend}
\addlegendentry{CP parabola}

\addlegendimage{draw=black, fill=mycolor26, area legend}
\addlegendentry{DR parabola}

\addlegendimage{draw=black, fill=mycolor27, area legend}
\addlegendentry{DL}

\addlegendimage{draw=black, fill=mycolor28, area legend}
\addlegendentry{NMF}

\addplot [forget plot] graphics [xmin=0.5, xmax=4.5, ymin=-1, ymax=22] {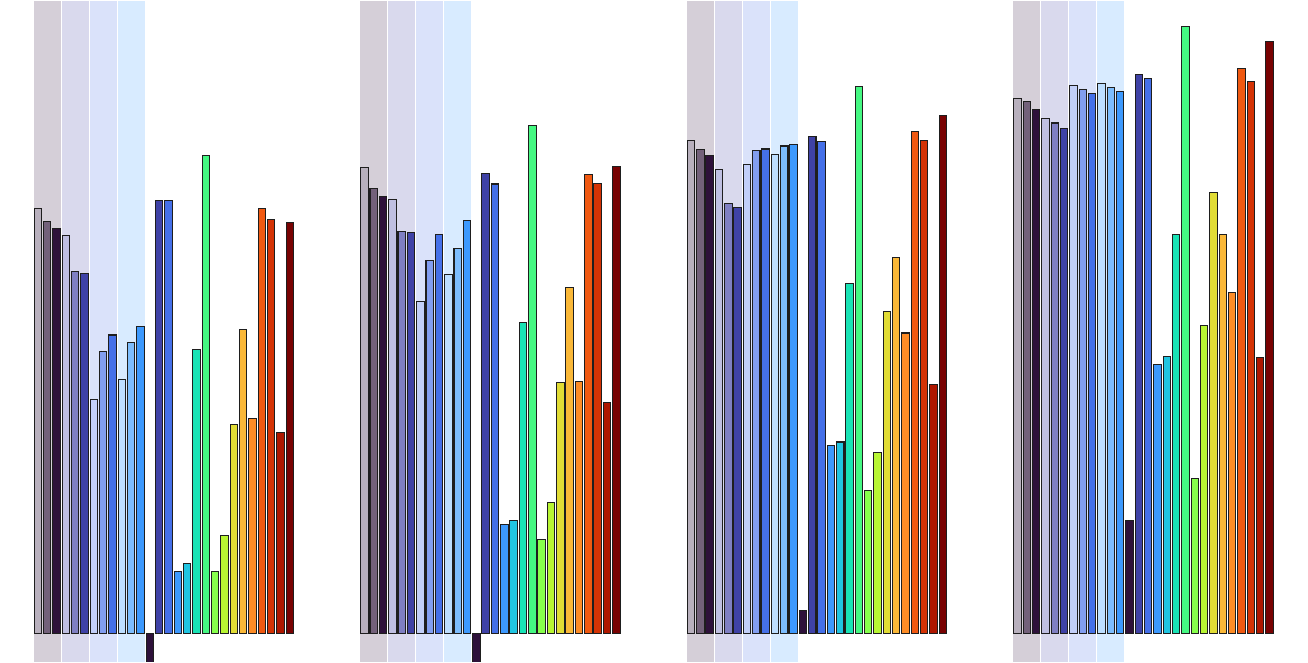};

\end{axis}
\end{tikzpicture}%
}
		\label{fig:survey.test:sdr}
	}
	\\[-0.1in]
	\subfloat[
		comparison in terms of PEMO-Q
	]
	{	
		\hspace{-0.8in}
		\adjustbox{scale=0.5}{% This file was created by matlab2tikz.
\definecolor{mycolor1}{rgb}{0.72998,0.69059,0.74406}
\definecolor{mycolor2}{rgb}{0.45997,0.38117,0.48811}
\definecolor{mycolor3}{rgb}{0.18995,0.07176,0.23217}
\definecolor{mycolor4}{rgb}{0.75123,0.75442,0.88469}
\definecolor{mycolor5}{rgb}{0.50246,0.50885,0.76937}
\definecolor{mycolor6}{rgb}{0.25369,0.26327,0.65406}
\definecolor{mycolor7}{rgb}{0.75897,0.81382,0.97109}
\definecolor{mycolor8}{rgb}{0.51794,0.62763,0.94219}
\definecolor{mycolor9}{rgb}{0.27691,0.44145,0.91328}
\definecolor{mycolor10}{rgb}{0.74809,0.86979,0.99899}
\definecolor{mycolor11}{rgb}{0.49618,0.73958,0.99798}
\definecolor{mycolor12}{rgb}{0.24427,0.60937,0.99697}
\definecolor{mycolor13}{rgb}{0.18995,0.07176,0.23217}
\definecolor{mycolor14}{rgb}{0.25369,0.26327,0.65406}
\definecolor{mycolor15}{rgb}{0.27691,0.44145,0.91328}
\definecolor{mycolor16}{rgb}{0.24427,0.60937,0.99697}
\definecolor{mycolor17}{rgb}{0.13278,0.77165,0.88580}
\definecolor{mycolor18}{rgb}{0.10342,0.89600,0.71500}
\definecolor{mycolor19}{rgb}{0.27597,0.97092,0.51653}
\definecolor{mycolor20}{rgb}{0.53255,0.99919,0.30581}
\definecolor{mycolor21}{rgb}{0.72596,0.96470,0.20640}
\definecolor{mycolor22}{rgb}{0.88331,0.86553,0.21719}
\definecolor{mycolor23}{rgb}{0.98000,0.73000,0.22161}
\definecolor{mycolor24}{rgb}{0.99297,0.55214,0.15417}
\definecolor{mycolor25}{rgb}{0.94084,0.35566,0.07031}
\definecolor{mycolor26}{rgb}{0.83926,0.20654,0.02305}
\definecolor{mycolor27}{rgb}{0.68602,0.09536,0.00481}
\definecolor{mycolor28}{rgb}{0.47960,0.01583,0.01055}
\begin{tikzpicture}[font=\large]

\begin{axis}[%
width=11in,
height=5in,
at={(0in,0in)},
scale only axis,
bar shift auto,
xmin=0.5,
xmax=4.5,
xtick={1,2,3,4},
xticklabels={{5},{7},{10},{15}},
xlabel style={font=\large},
xlabel={input SDR (dB)},
ymin=-4,
ymax=0,
ylabel style={font=\large},
ylabel={PEMO-Q ODG},
ytick={-4,-3.5,...,0},
yminorticks=true,
minor y tick num=5,
ymajorgrids,
yminorgrids,
axis background/.style={fill=white},
legend style={
	at={(1.03,1.0)},
	anchor=north west,
	legend cell align=left,
	align=left,
	/tikz/column 1/.style={
		column sep=4pt,
}}
]
]
\addlegendimage{draw=black, fill=mycolor1, area legend}
\addlegendentry{inp., $\lcoef= 0$}

\addlegendimage{draw=black, fill=mycolor2, area legend}
\addlegendentry{inp., $\lcoef = 10^{-5}$}

\addlegendimage{draw=black, fill=mycolor3, area legend}
\addlegendentry{inp., $\lcoef= 10^{-3}$}

\addlegendimage{draw=black, fill=mycolor4, area legend}
\addlegendentry{GLP, $\lcoef= 0$}

\addlegendimage{draw=black, fill=mycolor5, area legend}
\addlegendentry{GLP, $\lcoef = 10^{-5}$}

\addlegendimage{draw=black, fill=mycolor6, area legend}
\addlegendentry{GLP, $\lcoef= 10^{-3}$}

\addlegendimage{draw=black, fill=mycolor7, area legend}
\addlegendentry{dec., $\lcoef= 0$, $\lsig= 10$}

\addlegendimage{draw=black, fill=mycolor8, area legend}
\addlegendentry{dec., $\lcoef = 10^{-5}$, $\lsig= 10$}

\addlegendimage{draw=black, fill=mycolor9, area legend}
\addlegendentry{dec., $\lcoef= 10^{-3}$, $\lsig= 10$}

\addlegendimage{draw=black, fill=mycolor10, area legend}
\addlegendentry{dec., $\lcoef= 0$, $\lsig= \infty$}

\addlegendimage{draw=black, fill=mycolor11, area legend}
\addlegendentry{dec., $\lcoef = 10^{-5}$, $\lsig= \infty$}

\addlegendimage{draw=black, fill=mycolor12, area legend}
\addlegendentry{dec., $\lcoef= 10^{-3}$, $\lsig= \infty$}

\addlegendimage{draw=black, fill=mycolor13, area legend}
\addlegendentry{C-OMP}

\addlegendimage{draw=black, fill=mycolor14, area legend}
\addlegendentry{A-SPADE}

\addlegendimage{draw=black, fill=mycolor15, area legend}
\addlegendentry{S-SPADE}

\addlegendimage{draw=black, fill=mycolor16, area legend}
\addlegendentry{CP}

\addlegendimage{draw=black, fill=mycolor17, area legend}
\addlegendentry{DR}

\addlegendimage{draw=black, fill=mycolor18, area legend}
\addlegendentry{SS EW}

\addlegendimage{draw=black, fill=mycolor19, area legend}
\addlegendentry{SS PEW}

\addlegendimage{draw=black, fill=mycolor20, area legend}
\addlegendentry{CSL1}

\addlegendimage{draw=black, fill=mycolor21, area legend}
\addlegendentry{PCSL1}

\addlegendimage{draw=black, fill=mycolor22, area legend}
\addlegendentry{PWCSL1}

\addlegendimage{draw=black, fill=mycolor23, area legend}
\addlegendentry{reweighted CP}

\addlegendimage{draw=black, fill=mycolor24, area legend}
\addlegendentry{reweighted DR}

\addlegendimage{draw=black, fill=mycolor25, area legend}
\addlegendentry{CP parabola}

\addlegendimage{draw=black, fill=mycolor26, area legend}
\addlegendentry{DR parabola}

\addlegendimage{draw=black, fill=mycolor27, area legend}
\addlegendentry{DL}

\addlegendimage{draw=black, fill=mycolor28, area legend}
\addlegendentry{NMF}

\addplot [forget plot] graphics [xmin=0.5, xmax=4.5, ymin=-4, ymax=0] {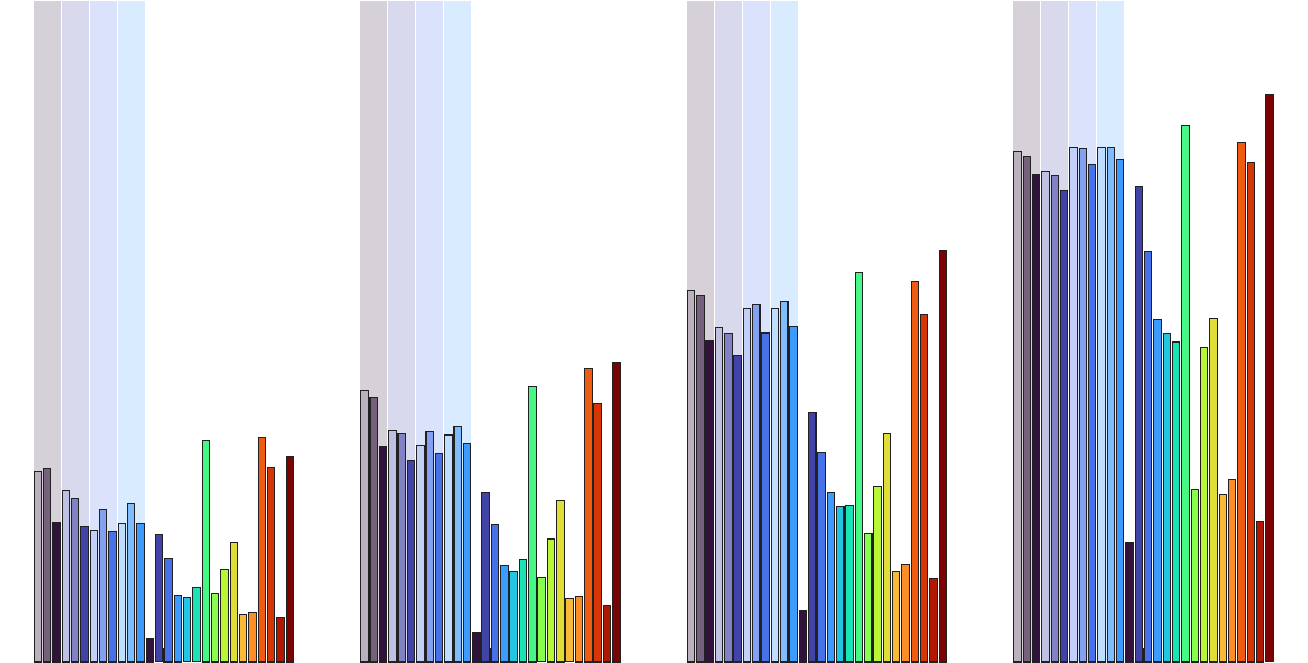};

\end{axis}
\end{tikzpicture}%
}
		\label{fig:survey.test:pemoq}
	}
	\\[-0.1in]
	\subfloat[
		comparison in terms of PEAQ
	]
	{
		\hspace{-0.8in}
		\adjustbox{scale=0.5}{% This file was created by matlab2tikz.
\definecolor{mycolor1}{rgb}{0.72998,0.69059,0.74406}
\definecolor{mycolor2}{rgb}{0.45997,0.38117,0.48811}
\definecolor{mycolor3}{rgb}{0.18995,0.07176,0.23217}
\definecolor{mycolor4}{rgb}{0.75123,0.75442,0.88469}
\definecolor{mycolor5}{rgb}{0.50246,0.50885,0.76937}
\definecolor{mycolor6}{rgb}{0.25369,0.26327,0.65406}
\definecolor{mycolor7}{rgb}{0.75897,0.81382,0.97109}
\definecolor{mycolor8}{rgb}{0.51794,0.62763,0.94219}
\definecolor{mycolor9}{rgb}{0.27691,0.44145,0.91328}
\definecolor{mycolor10}{rgb}{0.74809,0.86979,0.99899}
\definecolor{mycolor11}{rgb}{0.49618,0.73958,0.99798}
\definecolor{mycolor12}{rgb}{0.24427,0.60937,0.99697}
\definecolor{mycolor13}{rgb}{0.18995,0.07176,0.23217}
\definecolor{mycolor14}{rgb}{0.25369,0.26327,0.65406}
\definecolor{mycolor15}{rgb}{0.27691,0.44145,0.91328}
\definecolor{mycolor16}{rgb}{0.24427,0.60937,0.99697}
\definecolor{mycolor17}{rgb}{0.13278,0.77165,0.88580}
\definecolor{mycolor18}{rgb}{0.10342,0.89600,0.71500}
\definecolor{mycolor19}{rgb}{0.27597,0.97092,0.51653}
\definecolor{mycolor20}{rgb}{0.53255,0.99919,0.30581}
\definecolor{mycolor21}{rgb}{0.72596,0.96470,0.20640}
\definecolor{mycolor22}{rgb}{0.88331,0.86553,0.21719}
\definecolor{mycolor23}{rgb}{0.98000,0.73000,0.22161}
\definecolor{mycolor24}{rgb}{0.99297,0.55214,0.15417}
\definecolor{mycolor25}{rgb}{0.94084,0.35566,0.07031}
\definecolor{mycolor26}{rgb}{0.83926,0.20654,0.02305}
\definecolor{mycolor27}{rgb}{0.68602,0.09536,0.00481}
\definecolor{mycolor28}{rgb}{0.47960,0.01583,0.01055}
\begin{tikzpicture}[font=\large]

\begin{axis}[%
width=11in,
height=5in,
at={(0in,0in)},
scale only axis,
bar shift auto,
xmin=0.5,
xmax=4.5,
xtick={1,2,3,4},
xticklabels={{5},{7},{10},{15}},
xlabel style={font=\large},
xlabel={input SDR (dB)},
ymin=-4,
ymax=0,
ylabel style={font=\large},
ylabel={PEAQ ODG},
ytick={-4,-3.5,...,0},
yminorticks=true,
minor y tick num=5,
ymajorgrids,
yminorgrids,
axis background/.style={fill=white},
legend style={
	at={(1.03,1.0)},
	anchor=north west,
	legend cell align=left,
	align=left,
	/tikz/column 1/.style={
		column sep=4pt,
	}}
]
\addlegendimage{draw=black, fill=mycolor1, area legend}
\addlegendentry{inp., $\lcoef = 0$}

\addlegendimage{draw=black, fill=mycolor2, area legend}
\addlegendentry{inp., $\lcoef = 10^{-5}$}

\addlegendimage{draw=black, fill=mycolor3, area legend}
\addlegendentry{inp., $\lcoef = 10^{-3}$}

\addlegendimage{draw=black, fill=mycolor4, area legend}
\addlegendentry{GLP, $\lcoef = 0$}

\addlegendimage{draw=black, fill=mycolor5, area legend}
\addlegendentry{GLP, $\lcoef = 10^{-5}$}

\addlegendimage{draw=black, fill=mycolor6, area legend}
\addlegendentry{GLP, $\lcoef = 10^{-3}$}

\addlegendimage{draw=black, fill=mycolor7, area legend}
\addlegendentry{dec., $\lcoef = 0$, $\lsig = 10$}

\addlegendimage{draw=black, fill=mycolor8, area legend}
\addlegendentry{dec., $\lcoef = 10^{-5}$, $\lsig = 10$}

\addlegendimage{draw=black, fill=mycolor9, area legend}
\addlegendentry{dec., $\lcoef = 10^{-3}$, $\lsig = 10$}

\addlegendimage{draw=black, fill=mycolor10, area legend}
\addlegendentry{dec., $\lcoef = 0$, $\lsig = \infty$}

\addlegendimage{draw=black, fill=mycolor11, area legend}
\addlegendentry{dec., $\lcoef = 10^{-5}$, $\lsig = \infty$}

\addlegendimage{draw=black, fill=mycolor12, area legend}
\addlegendentry{dec., $\lcoef = 10^{-3}$, $\lsig = \infty$}

\addlegendimage{draw=black, fill=mycolor13, area legend}
\addlegendentry{C-OMP}

\addlegendimage{draw=black, fill=mycolor14, area legend}
\addlegendentry{A-SPADE}

\addlegendimage{draw=black, fill=mycolor15, area legend}
\addlegendentry{S-SPADE}

\addlegendimage{draw=black, fill=mycolor16, area legend}
\addlegendentry{CP}

\addlegendimage{draw=black, fill=mycolor17, area legend}
\addlegendentry{DR}

\addlegendimage{draw=black, fill=mycolor18, area legend}
\addlegendentry{SS EW}

\addlegendimage{draw=black, fill=mycolor19, area legend}
\addlegendentry{SS PEW}

\addlegendimage{draw=black, fill=mycolor20, area legend}
\addlegendentry{CSL1}

\addlegendimage{draw=black, fill=mycolor21, area legend}
\addlegendentry{PCSL1}

\addlegendimage{draw=black, fill=mycolor22, area legend}
\addlegendentry{PWCSL1}

\addlegendimage{draw=black, fill=mycolor23, area legend}
\addlegendentry{reweighted CP}

\addlegendimage{draw=black, fill=mycolor24, area legend}
\addlegendentry{reweighted DR}

\addlegendimage{draw=black, fill=mycolor25, area legend}
\addlegendentry{CP parabola}

\addlegendimage{draw=black, fill=mycolor26, area legend}
\addlegendentry{DR parabola}

\addlegendimage{draw=black, fill=mycolor27, area legend}
\addlegendentry{DL}

\addlegendimage{draw=black, fill=mycolor28, area legend}
\addlegendentry{NMF}

\addplot [forget plot] graphics [xmin=0.5, xmax=4.5, ymin=-4, ymax=0] {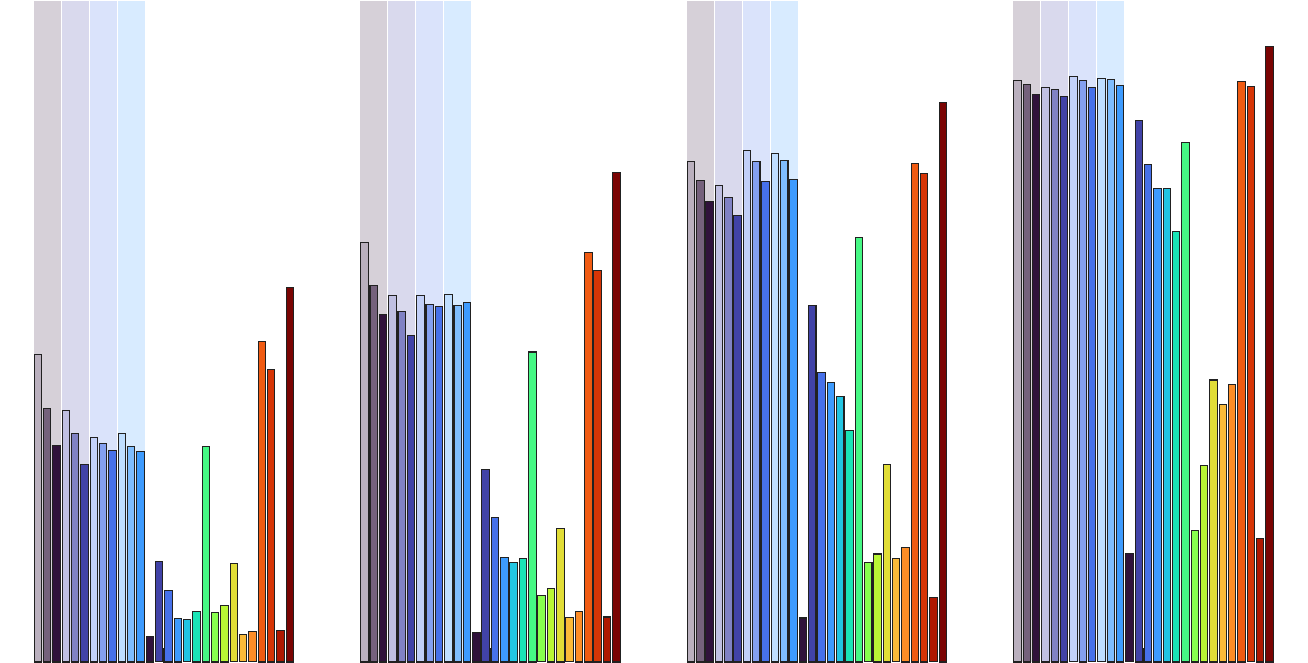};

\end{axis}
\end{tikzpicture}%
}
		\label{fig:survey.test:peaq}
	}
	\caption{%
		Comparison of the regularized AR model with the methods from the survey \cite{ZaviskaRajmicOzerovRencker2021:Declipping.Survey}.
		Results of the methods based on the AR model, treated in this article, are indicated with a~background color.
%		In the case of PEMO-Q and PEAQ,
%		results using the \qm{crossfade reliable} strategy from \cite{ZaviskaRajmicMokry2022:Declipping.crossfading} are shown using stacked bars.
%		For each algorithm which may produce signals inconsistent with the reliable samples, the cross-faded strategy is applied and the updated result is shown in lighter shade.
	}
	\label{fig:survey.test}
\end{figure*}

First, the results are evaluated using the SDR,
%(see equation \eqref{eq:sdr}),
shown in Fig.~\ref{fig:survey.test:sdr}.
More precisely, Fig.~\ref{fig:survey.test:sdr} shows the $\Delta$SDR values,
which are defined as the difference between the SDR of the recovered signal and the input SDR of the clipped signal. % (also denoted as the input SDR).

The effect of the regularization of the AR coefficients ($\lcoef$) in the case of inpainting and GLP appears to be rather negative in terms of the quality of the reconstruction.
On the other hand, it is beneficial in the case of declipping for input
\mbox{SDR\,$<$\,15\,dB}.
An important observation is that GLP does \emph{not} perform better than inpainting, and it outperforms declipping only for a~very low input SDR.
This suggests that if time-domain consistency is essential,
%a constraint,
%\todo{\textbf{PR:} jazykově zvláštní}
the declipping strategy represents a~better choice;
if computational complexity is the priority, then the inpainting strategy is sufficient.
For a~lower input SDR, the results are positively affected also by increasing $\lsig > 0$.

%\begin{figure*}
%	\centering
%	\subfloat[$\Delta$SDR]{\includegraphics[scale=0.55]{figures/SDR.pdf}} \\
%	\subfloat[PEMO-Q]{\includegraphics[scale=0.55]{figures/PEMOQ.pdf}} \\
%	\subfloat[PEAQ]{\includegraphics[scale=0.55]{figures/PEAQ.pdf}}
%	\caption{Comparison of the proposed methods with the methods from \cite{ZaviskaRajmicOzerovRencker2021:Declipping.Survey}.}
%	\todo[inline]{\textbf{Poznámka ke značení:} v obrázku $\lambda_1 = \lcoef, \lambda_2 = \lsig$}
%	\label{fig:survey.test}
%\end{figure*}

The overall outcome of the experiment is that
compared to the methods tested in the survey \cite{ZaviskaRajmicOzerovRencker2021:Declipping.Survey},
the regularized AR model scores among the best and offers an alternative to state-of-the-art methods.
However,
a~significant improvement of the state of the art is not observed.

Furthermore,
the regularization of the signal ($\lsig>0$) was expected to be clearly beneficial,
but it led to outperforming the other AR-based competitors (inpainting, GLP) for a~higher input SDR only.
For further discussion on the consistency of the declipped signals, see Appendix \ref{sec:consistency}.

\edt{%
%To conclude the experiment,
As the final part of this evaluation, %PR
Fig.~\ref{fig:wilcoxon} illustrates a~statistical significance test using the Wilcoxon signed rank test~\cite{Hollander1999}.
%\todo{OM: Odkaz na supplementary materials? Footnote?}
The test compares the results of all pairs of methods.
The right-tailed hypothesis test was configured such that rejection of the null hypothesis ($p\text{-value}$ smaller than $0.05$) suggests higher median of the first method from the couple under test.
As an example, the analysis reveals that all the proposed declipping cases outperform the GLP with model regularization ($\lcoef>0$).
The declipping case with $\lcoef=10^{-5}$ and $\lsig=\infty$ outperforms also the unmodified GLP, but with $p$-value close to the decision level.
In summary, the proposed methods significantly outperform 10 out of the 16 baselines.
Statistical tests for other values of the input SDR are presented at the accompanying GitHub page (see the link in Sec.~\ref{ssec:software}).
}

\begin{figure}
	\adjustbox{scale=0.55}{\input{figures/wilcoxon.tex}}
	\caption{%
		\edt{%
			Illustration of testing the statistical hypothesis on the $\Delta$SDR results for the case of input SDR 10\,dB
			(i.e., third group of columns in Fig.~\ref{fig:survey.test:sdr}).
			Each entry of the matrix represents a~comparison of the method on the vertical axis to the method on the horizontal axis.
			Entries with $p\text{-value}<0.05$ suggest superiority of the corresponding method on the vertical axis and are colored in shades of blue.
			Failure to reject the null hypothesis (i.e., possible equality of the medians) is colored in red.
		}
	}
	\label{fig:wilcoxon}
\end{figure}

%
%\todo{PR: Zbytek odstavce přesunout do appendixu, drobně rozšířit a přidat nový graf SDR versus consistency. (použijeme scatter plot, jiné barvy než ve sloupečkách, přiměřeně oříznuté osy)}
%A hypothesis is that
%although consistency in the time domain (satisfying the clipping conditions) can be approached with $\lsig>0$ and even guaranteed in the case $\lsig = \infty$,
%we are forced to numerically solve subproblem \eqref{eq:ACS2:x}, which carries risks (only approximate solution due to the limited number of iterations, possible insufficient bias of the declipped samples).

%\todo[inline]{PR: Pomohli bychom si, kdybychom úplně vynechali post-processing? Taky by se to pročistilo, takto je to pro čtenáře hodně variant. \\ OM: Ještě to probereme, ale klidně pročišťme. \\ *: Ano, uděláme to tak. Upravit grafy a text!}

%\subsection{Subjective test}
%\todo[inline]{\textbf{OM:} chtělo by to poslechový test...}

\subsection{\edt{Speech declipping}}
\label{ssec:speech}

% \todo{OM: určitě jde vyhodit: ViSQOL, footnote k SS\,PEW.\\ PR: přikláním se k tomu je nechat}

\edt{%
To strengthen the generality of the proposed framework,
an extension of the previous comparison focuses on declipping of speech.
This experiment uses a~subset of 20 signals from the dataset \cite{ValentiniBotinhao2017:Noisy.speech.database} (2~speakers, 1 male and 1 female, 10 signals each).
% while keeping the file names:
% 10 files with speaker p232 (male),
% 10 files with speaker p257 (female).
The original sounds represent clean speech excerpts with the length ranging from 1.6 to 3.8 seconds, with no significant portions of silence, sampled at 48~kHz.
%The clipping levels are defined by input SDR.
%\todo{PR: Tuto větu už bych tam nedával.}
}

\edt{%
The hyperparameters of the reconstruction were adapted to speech, namely the AR model order $p = 256$ and frame length $w=2048$ samples (approx.\ 43\,ms) were used.
Three baselines were selected from the methods tested in the survey \cite{ZaviskaRajmicOzerovRencker2021:Declipping.Survey} and applied with the frame length of $w=2048$ samples.
}

\edt{%
For speech-specific evaluation,
the Short-Time Objective Intelligibility (STOI) is employed,
ranging from $-1$ to $1$ and
providing an intelligibility-based comparison of a test signal to a reference signal \cite{Taal2010:STOI.ICASSP,Taal2011:STOI.TASLP}.
Besides,
Virtual Speech Quality Objective Listener (ViSQOL)
is used to assess psychoacoustical similarity of speech signals.
This model uses two scores:
%to assess the similarity of the test signal and the reference:
the mean opinion score (MOS) and the neurogram similarity index measure (NSIM) \cite{Hines2015:ViSQOL,Chinen2020:ViSQOL.v3, Hines2012:Speech.intelligibility.prediction.NSIM}.
}

\begin{figure}[ht]
	\centering
	\subfloat[
		comparison in terms of $\Delta$SDR, i.e.\ the improvement of SDR over the clipped signal
	]{
		\hspace{-0.4in}
		\adjustbox{scale=0.4}{% This file was created by matlab2tikz.
\definecolor{mycolor1}{rgb}{0.72998,0.69059,0.74406}
\definecolor{mycolor2}{rgb}{0.45997,0.38117,0.48811}
\definecolor{mycolor3}{rgb}{0.18995,0.07176,0.23217}
\definecolor{mycolor4}{rgb}{0.75123,0.75442,0.88469}
\definecolor{mycolor5}{rgb}{0.50246,0.50885,0.76937}
\definecolor{mycolor6}{rgb}{0.25369,0.26327,0.65406}
\definecolor{mycolor7}{rgb}{0.75897,0.81382,0.97109}
\definecolor{mycolor8}{rgb}{0.51794,0.62763,0.94219}
\definecolor{mycolor9}{rgb}{0.27691,0.44145,0.91328}
\definecolor{mycolor10}{rgb}{0.74809,0.86979,0.99899}
\definecolor{mycolor11}{rgb}{0.49618,0.73958,0.99798}
\definecolor{mycolor12}{rgb}{0.24427,0.60937,0.99697}
%\definecolor{mycolor13}{rgb}{0.18995,0.07176,0.23217}
\definecolor{mycolor14}{rgb}{0.25369,0.26327,0.65406}
%\definecolor{mycolor15}{rgb}{0.27691,0.44145,0.91328}
%\definecolor{mycolor16}{rgb}{0.24427,0.60937,0.99697}
%\definecolor{mycolor17}{rgb}{0.13278,0.77165,0.88580}
%\definecolor{mycolor18}{rgb}{0.10342,0.89600,0.71500}
\definecolor{mycolor19}{rgb}{0.27597,0.97092,0.51653}
\definecolor{mycolor20}{rgb}{0.53255,0.99919,0.30581}
%\definecolor{mycolor21}{rgb}{0.72596,0.96470,0.20640}
%\definecolor{mycolor22}{rgb}{0.88331,0.86553,0.21719}
%\definecolor{mycolor23}{rgb}{0.98000,0.73000,0.22161}
%\definecolor{mycolor24}{rgb}{0.99297,0.55214,0.15417}
%\definecolor{mycolor25}{rgb}{0.94084,0.35566,0.07031}
%\definecolor{mycolor26}{rgb}{0.83926,0.20654,0.02305}
%\definecolor{mycolor27}{rgb}{0.68602,0.09536,0.00481}
%\definecolor{mycolor28}{rgb}{0.47960,0.01583,0.01055}
%
\begin{tikzpicture}[font=\large]

\begin{axis}[%
width=5.5in,
height=3.2in,
at={(0in,0in)},
scale only axis,
bar shift auto,
xmin=0.5,
xmax=4.5,
xtick={1,2,3,4},
xticklabels={{5},{7},{10},{15}},
xlabel style={font=\large},
xlabel={input SDR (dB)},
ymin=-1,
ymax=10,
ylabel style={font=\large},
ylabel={$\Delta$SDR (dB)},
ytick={-2,0,2,4,6,8,10},
yminorticks=true,
minor y tick num=5,
ymajorgrids,
yminorgrids,
axis background/.style={fill=white},
legend style={
	at={(1.03,1.0)},
	anchor=north west,
	legend cell align=left,
	align=left,
	/tikz/column 1/.style={
		column sep=4pt,
}}
]
]
\addlegendimage{draw=black, fill=mycolor1, area legend}
\addlegendentry{inp., $\lcoef= 0$}

\addlegendimage{draw=black, fill=mycolor2, area legend}
\addlegendentry{inp., $\lcoef = 10^{-5}$}

\addlegendimage{draw=black, fill=mycolor3, area legend}
\addlegendentry{inp., $\lcoef= 10^{-3}$}

\addlegendimage{draw=black, fill=mycolor4, area legend}
\addlegendentry{GLP, $\lcoef= 0$}

\addlegendimage{draw=black, fill=mycolor5, area legend}
\addlegendentry{GLP, $\lcoef = 10^{-5}$}

\addlegendimage{draw=black, fill=mycolor6, area legend}
\addlegendentry{GLP, $\lcoef= 10^{-3}$}

\addlegendimage{draw=black, fill=mycolor7, area legend}
\addlegendentry{dec., $\lcoef= 0$, $\lsig= 10$}

\addlegendimage{draw=black, fill=mycolor8, area legend}
\addlegendentry{dec., $\lcoef = 10^{-5}$, $\lsig= 10$}

\addlegendimage{draw=black, fill=mycolor9, area legend}
\addlegendentry{dec., $\lcoef= 10^{-3}$, $\lsig= 10$}

\addlegendimage{draw=black, fill=mycolor10, area legend}
\addlegendentry{dec., $\lcoef= 0$, $\lsig= \infty$}

\addlegendimage{draw=black, fill=mycolor11, area legend}
\addlegendentry{dec., $\lcoef = 10^{-5}$, $\lsig= \infty$}

\addlegendimage{draw=black, fill=mycolor12, area legend}
\addlegendentry{dec., $\lcoef= 10^{-3}$, $\lsig= \infty$}

%\addlegendimage{draw=black, fill=mycolor13, area legend}
%\addlegendentry{C-OMP}

\addlegendimage{draw=black, fill=mycolor14, area legend}
\addlegendentry{A-SPADE}

%\addlegendimage{draw=black, fill=mycolor15, area legend}
%\addlegendentry{S-SPADE}
%
%\addlegendimage{draw=black, fill=mycolor16, area legend}
%\addlegendentry{CP}
%
%\addlegendimage{draw=black, fill=mycolor17, area legend}
%\addlegendentry{DR}
%
%\addlegendimage{draw=black, fill=mycolor18, area legend}
%\addlegendentry{SS EW}

\addlegendimage{draw=black, fill=mycolor19, area legend}
\addlegendentry{SS PEW}

\addlegendimage{draw=black, fill=mycolor20, area legend}
\addlegendentry{CSL1}

%\addlegendimage{draw=black, fill=mycolor21, area legend}
%\addlegendentry{PCSL1}
%
%\addlegendimage{draw=black, fill=mycolor22, area legend}
%\addlegendentry{PWCSL1}
%
%\addlegendimage{draw=black, fill=mycolor23, area legend}
%\addlegendentry{reweighted CP}
%
%\addlegendimage{draw=black, fill=mycolor24, area legend}
%\addlegendentry{reweighted DR}
%
%\addlegendimage{draw=black, fill=mycolor25, area legend}
%\addlegendentry{CP parabola}
%
%\addlegendimage{draw=black, fill=mycolor26, area legend}
%\addlegendentry{DR parabola}
%
%\addlegendimage{draw=black, fill=mycolor27, area legend}
%\addlegendentry{DL}
%
%\addlegendimage{draw=black, fill=mycolor28, area legend}
%\addlegendentry{NMF}

\addplot [forget plot] graphics [xmin=0.5, xmax=4.5, ymin=-1, ymax=10] {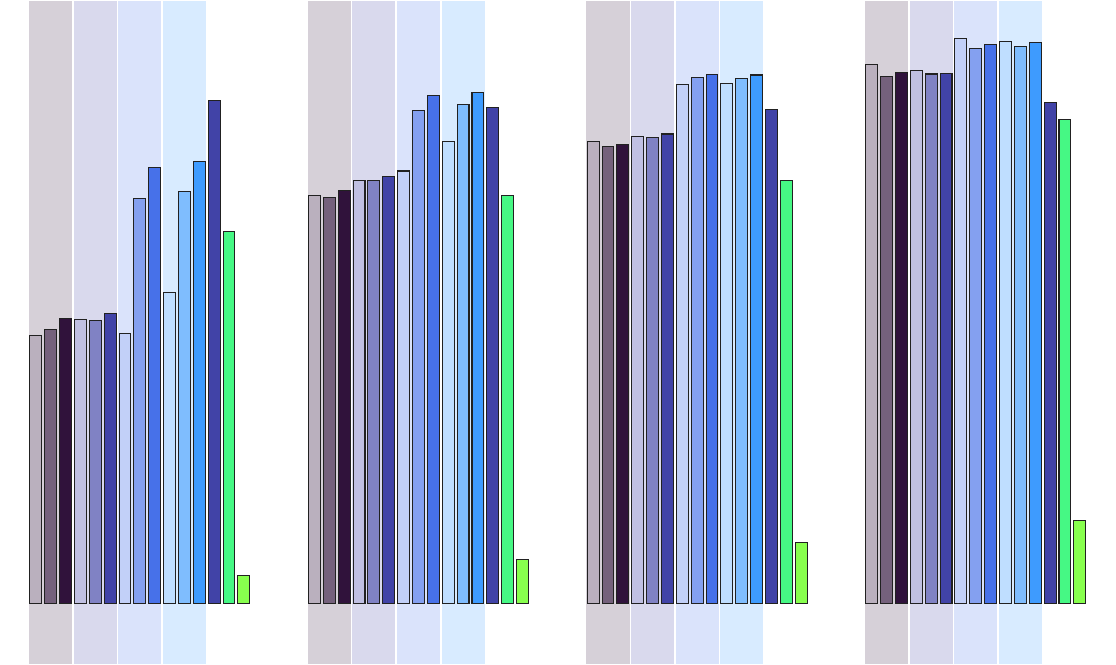};

\end{axis}
\end{tikzpicture}%
}%
		\label{fig:speech.test:sdr}
	}
	\\
	\subfloat[
		comparison in terms of STOI (note the cropped y-axis)
	]{
		\hspace{-0.4in}
		\adjustbox{scale=0.4}{% This file was created by matlab2tikz.
\definecolor{mycolor1}{rgb}{0.72998,0.69059,0.74406}
\definecolor{mycolor2}{rgb}{0.45997,0.38117,0.48811}
\definecolor{mycolor3}{rgb}{0.18995,0.07176,0.23217}
\definecolor{mycolor4}{rgb}{0.75123,0.75442,0.88469}
\definecolor{mycolor5}{rgb}{0.50246,0.50885,0.76937}
\definecolor{mycolor6}{rgb}{0.25369,0.26327,0.65406}
\definecolor{mycolor7}{rgb}{0.75897,0.81382,0.97109}
\definecolor{mycolor8}{rgb}{0.51794,0.62763,0.94219}
\definecolor{mycolor9}{rgb}{0.27691,0.44145,0.91328}
\definecolor{mycolor10}{rgb}{0.74809,0.86979,0.99899}
\definecolor{mycolor11}{rgb}{0.49618,0.73958,0.99798}
\definecolor{mycolor12}{rgb}{0.24427,0.60937,0.99697}
%\definecolor{mycolor13}{rgb}{0.18995,0.07176,0.23217}
\definecolor{mycolor14}{rgb}{0.25369,0.26327,0.65406}
%\definecolor{mycolor15}{rgb}{0.27691,0.44145,0.91328}
%\definecolor{mycolor16}{rgb}{0.24427,0.60937,0.99697}
%\definecolor{mycolor17}{rgb}{0.13278,0.77165,0.88580}
%\definecolor{mycolor18}{rgb}{0.10342,0.89600,0.71500}
\definecolor{mycolor19}{rgb}{0.27597,0.97092,0.51653}
\definecolor{mycolor20}{rgb}{0.53255,0.99919,0.30581}
%\definecolor{mycolor21}{rgb}{0.72596,0.96470,0.20640}
%\definecolor{mycolor22}{rgb}{0.88331,0.86553,0.21719}
%\definecolor{mycolor23}{rgb}{0.98000,0.73000,0.22161}
%\definecolor{mycolor24}{rgb}{0.99297,0.55214,0.15417}
%\definecolor{mycolor25}{rgb}{0.94084,0.35566,0.07031}
%\definecolor{mycolor26}{rgb}{0.83926,0.20654,0.02305}
%\definecolor{mycolor27}{rgb}{0.68602,0.09536,0.00481}
%\definecolor{mycolor28}{rgb}{0.47960,0.01583,0.01055}
%
\begin{tikzpicture}[font=\large]

\begin{axis}[%
width=5.5in,
height=3.2in,
at={(0in,0in)},
scale only axis,
bar shift auto,
xmin=0.5,
xmax=4.5,
xtick={1,2,3,4},
xticklabels={{5},{7},{10},{15}},
xlabel style={font=\large},
xlabel={input SDR (dB)},
ymin=0.8,
ymax=1.0,
ylabel style={font=\large},
ylabel={STOI},
ytick={0.80,0.85,0.90,0.95,1.00},
yminorticks=true,
minor y tick num=5,
ymajorgrids,
yminorgrids,
axis background/.style={fill=white},
legend style={
	at={(1.03,1.0)},
	anchor=north west,
	legend cell align=left,
	align=left,
	/tikz/column 1/.style={
		column sep=4pt,
}}
]
]
\addlegendimage{draw=black, fill=mycolor1, area legend}
\addlegendentry{inp., $\lcoef= 0$}

\addlegendimage{draw=black, fill=mycolor2, area legend}
\addlegendentry{inp., $\lcoef = 10^{-5}$}

\addlegendimage{draw=black, fill=mycolor3, area legend}
\addlegendentry{inp., $\lcoef= 10^{-3}$}

\addlegendimage{draw=black, fill=mycolor4, area legend}
\addlegendentry{GLP, $\lcoef= 0$}

\addlegendimage{draw=black, fill=mycolor5, area legend}
\addlegendentry{GLP, $\lcoef = 10^{-5}$}

\addlegendimage{draw=black, fill=mycolor6, area legend}
\addlegendentry{GLP, $\lcoef= 10^{-3}$}

\addlegendimage{draw=black, fill=mycolor7, area legend}
\addlegendentry{dec., $\lcoef= 0$, $\lsig= 10$}

\addlegendimage{draw=black, fill=mycolor8, area legend}
\addlegendentry{dec., $\lcoef = 10^{-5}$, $\lsig= 10$}

\addlegendimage{draw=black, fill=mycolor9, area legend}
\addlegendentry{dec., $\lcoef= 10^{-3}$, $\lsig= 10$}

\addlegendimage{draw=black, fill=mycolor10, area legend}
\addlegendentry{dec., $\lcoef= 0$, $\lsig= \infty$}

\addlegendimage{draw=black, fill=mycolor11, area legend}
\addlegendentry{dec., $\lcoef = 10^{-5}$, $\lsig= \infty$}

\addlegendimage{draw=black, fill=mycolor12, area legend}
\addlegendentry{dec., $\lcoef= 10^{-3}$, $\lsig= \infty$}

%\addlegendimage{draw=black, fill=mycolor13, area legend}
%\addlegendentry{C-OMP}

\addlegendimage{draw=black, fill=mycolor14, area legend}
\addlegendentry{A-SPADE}

%\addlegendimage{draw=black, fill=mycolor15, area legend}
%\addlegendentry{S-SPADE}
%
%\addlegendimage{draw=black, fill=mycolor16, area legend}
%\addlegendentry{CP}
%
%\addlegendimage{draw=black, fill=mycolor17, area legend}
%\addlegendentry{DR}
%
%\addlegendimage{draw=black, fill=mycolor18, area legend}
%\addlegendentry{SS EW}

\addlegendimage{draw=black, fill=mycolor19, area legend}
\addlegendentry{SS PEW}

\addlegendimage{draw=black, fill=mycolor20, area legend}
\addlegendentry{CSL1}

%\addlegendimage{draw=black, fill=mycolor21, area legend}
%\addlegendentry{PCSL1}
%
%\addlegendimage{draw=black, fill=mycolor22, area legend}
%\addlegendentry{PWCSL1}
%
%\addlegendimage{draw=black, fill=mycolor23, area legend}
%\addlegendentry{reweighted CP}
%
%\addlegendimage{draw=black, fill=mycolor24, area legend}
%\addlegendentry{reweighted DR}
%
%\addlegendimage{draw=black, fill=mycolor25, area legend}
%\addlegendentry{CP parabola}
%
%\addlegendimage{draw=black, fill=mycolor26, area legend}
%\addlegendentry{DR parabola}
%
%\addlegendimage{draw=black, fill=mycolor27, area legend}
%\addlegendentry{DL}
%
%\addlegendimage{draw=black, fill=mycolor28, area legend}
%\addlegendentry{NMF}

\addplot [forget plot] graphics [xmin=0.5, xmax=4.5, ymin=0.8, ymax=1.0] {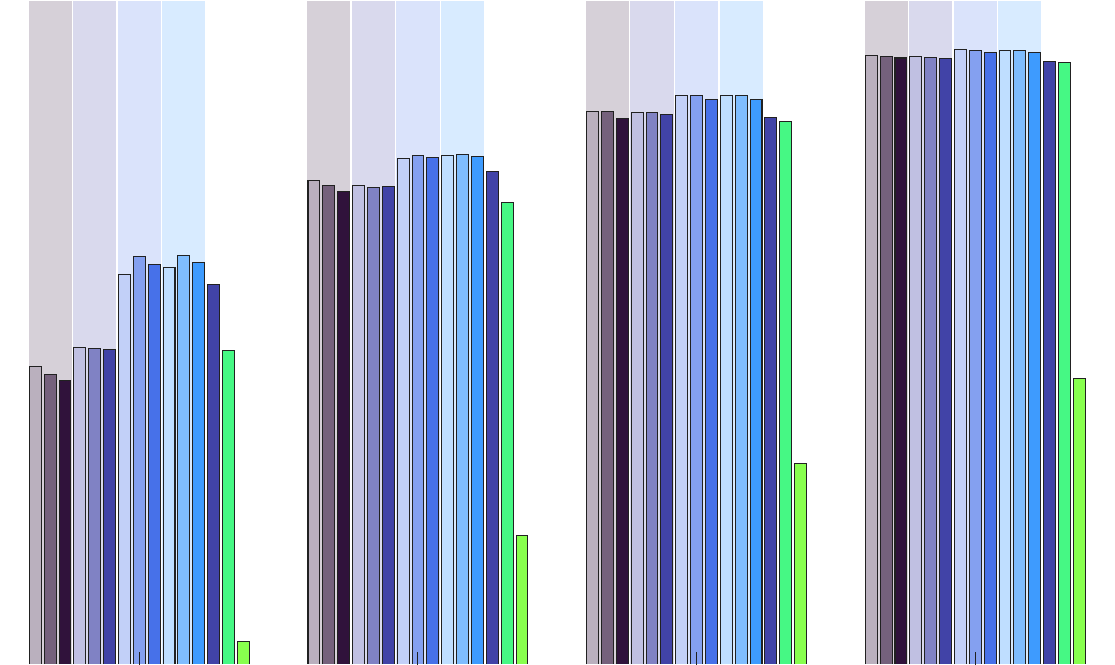};

\end{axis}
\end{tikzpicture}%
}%
		\label{fig:speech.test:stoi}
	}
	\\
	\caption{%
		\edt{%
		The proposed regularized AR model is compared with selected methods from the survey \cite{ZaviskaRajmicOzerovRencker2021:Declipping.Survey},
		here on a~speech dataset.
		The color-coding follows the experiment on musical signals in Sec.\ \ref{ssec:comparison.survey} and Fig.\ \ref{fig:survey.test}.
		}
	}
	\label{fig:speech.test}
\end{figure}

\edt{%
The results in terms of $\Delta$SDR and STOI are shown in Fig.~\ref{fig:speech.test}.
Regarding $\Delta$SDR and the proposed methods,
both consistent and inconsistent declipping with ACS outperform the inpainting and GLP cases.
Positive effect of the AR model regularization ($\lcoef > 0$) is also pronounced.
Importantly, the proposed AR-based declipping also outperforms the baselines with the only exception of A-SPADE at input SDR of 5\,dB.
It is worth noting that SS\,PEW was selected as the best performing from Sec.\ \ref{ssec:comparison.survey};
however,
it does not reach the results of \mbox{A-SPADE} in the case of speech.\!%
\footnote{%
	This can be attributed to the fact that SS\,PEW exploits the persistence properties of sounds, which is significant in music but only concerns voiced segments of speech in a~limited manner.
}
Regarding STOI,
the differences are subtle and the effect of $\lcoef$ appears insignificant.
On the other hand,
the proposed methods outperform the baselines for all clipping levels.
Equivalent outcomes to those of STOI stem from the MOS and NSIM metrics, which are not displayed.
}

%\vspace{-\baselineskip}%
\subsection{Audio dequantization}%
\label{ssec:dequantization}

The audio dequantization task can also be formalized within the framework of regularized AR modeling in the form of problem \eqref{eq:AAR.declipping}.
The difference to audio declipping lies only in the consistency set:
$\setdec$ is now replaced by $\setdeq$ as defined in \eqref{eq:Gamma.dequant.definition}.
%As in the case of declipping,
The same as above,  %PR
$\lsig = \infty$ corresponds to the consistent case and $0 < \lsig < \infty$ to the inconsistent case.

%\todo[inline]{OM: Dopsat, jak to vychází celkově? PR: Jakože radši se vyhnout nepříjemnému numerickému srovnání? :) *: Radši se vyhnout, je to hrůza.}
%\todo[inline]{OM: Dopsat aspoň hodnoty SNR/ODG pro signály v obrázku? *: Radši se vyhnout, je to hrůza.}
%\todo[inline]{OM: Dopsat nějaké další nastavení, např.\ že je to 2.\ iterace? *: Je to moc detail.}
%\todo[inline]{Užitečné citace (sem nebo do úvodu): \\
%	Řeč, řídkost: \cite{BrauerZhaoLorenzFingscheidt2019:Dequantization_speech_signals} \\
%	Reimplementace + další experimenty: \cite{ZaviskaRajmic2020:Dequantization} \\
%	Víc metod: \cite{ZaviskaRajmicMokry2021:Audio.dequantization.ICASSP} (odsud je S-SPADQ) \\
%	Learning: \cite{BrauerZhaoLorenzFingscheidt2019:Dequantization_speech_signals} \\
%	Další řídkost: \cite{RenckerBachWangPlumbley2018:Fast.iterative.shrinkage.declip.dequant-iTwist18, RenckerBachWangPlumbley2018:Sparse.recovery.dictionary.learning} \\
%	Vzhledem k ilustrační povaze této části bych se možná omezil jenom na \cite{ZaviskaRajmicMokry2021:Audio.dequantization.ICASSP} kvůli S-SPADQ.
%}

Fig.\ \ref{fig:dequant} shows an illustrative reconstruction of a quantized piano recording.
The clean recording is taken from the dataset described in Sec.\ \ref{ssec:comparison.survey} and the uniform quantization from \eqref{eq:uniform_quantization} is used with 5 bits per sample, corresponding to a total of 32 quantization levels with a step of $\Delta = 2^{-5+1} = 0.0625$.

The plot shows a detail of the waveform of the clean signal, the reference reconstruction by the sparsity-based S-SPADQ algorithm \cite{ZaviskaRajmicMokry2021:Audio.dequantization.ICASSP} and, finally, the proposed regularized AR approach, with different regularization strengths.

%\todo{PR: Tady se obávám, že obrázek bude muset ukazovat kratší úsek v čase, nebo jej naopak roztáhnout přes oba sloupce.}
\begin{figure*}[ht]
	\centering
	\adjustbox{width=\linewidth}{\input{figures/dequant.tex}}
	\vspace{-1.5\baselineskip}%  %pomenil jsem to tady, aby fungoval krizovy odkaz
	\caption{Audio dequantization experiment, piano, 5-bit quantization. Each sample is constrained by the nearest decision levels, shown by the gray lines.}
	\label{fig:dequant}
%	            quantized,  SDR = 18.98, PEMO-Q ODG = -3.66, PEAQ ODG = -3.90
%	              S_SPADQ, dSDR =  8.14, PEMO-Q ODG = -3.31, PEAQ ODG = -3.84
%	         Janssen_cons, dSDR =  4.67, PEMO-Q ODG = -3.60, PEAQ ODG = -3.91
%	       Janssen_incons, dSDR =  4.70, PEMO-Q ODG = -3.59, PEAQ ODG = -3.91 lsig = 10
%          Janssen_incons, dSDR =  4.74, PEMO-Q ODG = -3.54, PEAQ ODG = -3.90 lsig = 1
%          Janssen_incons, dSDR =  4.36, PEMO-Q ODG = -3.46, PEAQ ODG = -3.80 lsig = 0.1
%	  Janssen_sparse_cons, dSDR =  1.77, PEMO-Q ODG = -3.87, PEAQ ODG = -3.88
%	Janssen_sparse_incons, dSDR =  1.66, PEMO-Q ODG = -3.87, PEAQ ODG = -3.88 lsig = 10
%   Janssen_sparse_incons, dSDR =  0.93, PEMO-Q ODG = -3.84, PEAQ ODG = -3.87 lsig = 1
%   Janssen_sparse_incons, dSDR = -1.63, PEMO-Q ODG = -3.79, PEAQ ODG = -3.43 lsig = 0.1
\end{figure*}

The example shows a variety of reconstructions depending on both the signal consistency (values of $\lsig$) and AR model regularization (values of $\lcoef$).
Interestingly,
from the two possibly inconsistent solutions ($\lsig=0.1$), we observe actual breaking of the consistency constraints only in the case of the regularized AR model ($\lcoef=0.1$).

\subsection{Computational complexity}
\label{ssec:complexity}

Regarding the computational demands of the proposed methods,
several insights can be reported:
\begin{itemize}
	\item
	The most demanding is the iterative solution of sub-problems \eqref{eq:ACS2:a} (if $\lcoef>0$) and \eqref{eq:ACS2:x}
	(see also Sec.~\ref{sec:algorithms.for.subproblems}).
	From \eqref{eq:matrices} and \eqref{eq:prox.error}, we observe that the computation is dominated by matrix manipulations,
	where the dimensions grow with both the signal/window length $N$ and model order $p$.
%	\todo{PR: Napsat, že rozměry matic rostou lineárně, ale inverze je O($N^3$), takže špatné?! \\ OM: doplněna jedna věta \\ PR: ok}
	More precisely, the complexity of matrix inversion (needed every outer iteration) is O($N^3$) in the signal update \eqref{eq:ACS2:x} and O($p^3$) in the model update \eqref{eq:ACS2:a}.
	\item
	In practical cases it holds $p \ll N$, therefore the model-related sub-problem \eqref{eq:ACS2:a} is numerically less demanding compared to the signal-related sub-problem \eqref{eq:ACS2:x}.
	\item
	In the case of inpainting and GLP, the signal estimation is done using matrix inversion,
	where the matrix size grows with the signal/window length and the number of missing/clipped samples (which is in an inverse relationship with the clipping threshold $\tc$ and with the input SDR).
	\item
	On the other hand, the computational load of the signal estimation \eqref{eq:ACS2:x} in the case of declipping with ACS does not depend on the number of degraded samples.
\end{itemize}

\begin{figure}[ht]
	\centering
	\hspace{-0.4in}
	\adjustbox{scale=0.485}{% This file was created by matlab2tikz.
\definecolor{mycolor1}{rgb}{0.72998,0.69059,0.74406}
\definecolor{mycolor2}{rgb}{0.45997,0.38117,0.48811}
\definecolor{mycolor3}{rgb}{0.18995,0.07176,0.23217}
\definecolor{mycolor4}{rgb}{0.75123,0.75442,0.88469}
\definecolor{mycolor5}{rgb}{0.50246,0.50885,0.76937}
\definecolor{mycolor6}{rgb}{0.25369,0.26327,0.65406}
\definecolor{mycolor7}{rgb}{0.75897,0.81382,0.97109}
\definecolor{mycolor8}{rgb}{0.51794,0.62763,0.94219}
\definecolor{mycolor9}{rgb}{0.27691,0.44145,0.91328}
\definecolor{mycolor10}{rgb}{0.74809,0.86979,0.99899}
\definecolor{mycolor11}{rgb}{0.49618,0.73958,0.99798}
\definecolor{mycolor12}{rgb}{0.24427,0.60937,0.99697}
\begin{tikzpicture}[font=\large]

\begin{axis}[%
width=5.5in,
height=3.2in,
at={(0in,0in)},
scale only axis,
bar shift auto,
xmin=0.5,
xmax=4.5,
xtick={1,2,3,4},
xticklabels={{5},{7},{10},{15}},
xlabel style={font=\large},
xlabel={input SDR (dB)},
ymin=1,
ymax=600,
ymode=log,
ylabel style={font=\large},
ylabel={elapsed time per second of audio (s)},
yminorticks=true,
minor y tick num=10,
ymajorgrids,
yminorgrids,
axis background/.style={fill=white},
legend columns=3,
transpose legend,
legend style={
	at={(0.5,1.05)},
	anchor=south,
	legend cell align=left,
	align=left,
	/tikz/column 2/.style={
		column sep=4pt,
	},
	/tikz/column 4/.style={
		column sep=4pt,
	},
	/tikz/column 6/.style={
		column sep=4pt,
	},
	/tikz/column 1/.style={
		column sep=2pt,
	},
	/tikz/column 3/.style={
		column sep=2pt,
	},
	/tikz/column 5/.style={
		column sep=2pt,
	},
	/tikz/column 7/.style={
		column sep=2pt,
	},
}
]
\addlegendimage{draw=black, fill=mycolor1, area legend}
\addlegendentry{inp., $\lcoef= 0$}

\addlegendimage{draw=black, fill=mycolor2, area legend}
\addlegendentry{inp., $\lcoef = 10^{-5}$}

\addlegendimage{draw=black, fill=mycolor3, area legend}
\addlegendentry{inp., $\lcoef= 10^{-3}$}

\addlegendimage{draw=black, fill=mycolor4, area legend}
\addlegendentry{GLP, $\lcoef= 0$}

\addlegendimage{draw=black, fill=mycolor5, area legend}
\addlegendentry{GLP, $\lcoef = 10^{-5}$}

\addlegendimage{draw=black, fill=mycolor6, area legend}
\addlegendentry{GLP, $\lcoef= 10^{-3}$}

\addlegendimage{draw=black, fill=mycolor7, area legend}
\addlegendentry{dec., $\lcoef= 0$, $\lsig= 10$}

\addlegendimage{draw=black, fill=mycolor8, area legend}
\addlegendentry{dec., $\lcoef = 10^{-5}$, $\lsig= 10$}

\addlegendimage{draw=black, fill=mycolor9, area legend}
\addlegendentry{dec., $\lcoef= 10^{-3}$, $\lsig= 10$}

\addlegendimage{draw=black, fill=mycolor10, area legend}
\addlegendentry{dec., $\lcoef= 0$, $\lsig= \infty$}

\addlegendimage{draw=black, fill=mycolor11, area legend}
\addlegendentry{dec., $\lcoef = 10^{-5}$, $\lsig= \infty$}

\addlegendimage{draw=black, fill=mycolor12, area legend}
\addlegendentry{dec., $\lcoef= 10^{-3}$, $\lsig= \infty$}

\addplot [forget plot] graphics [xmin=0.5, xmax=4.5, ymin=1, ymax=600] {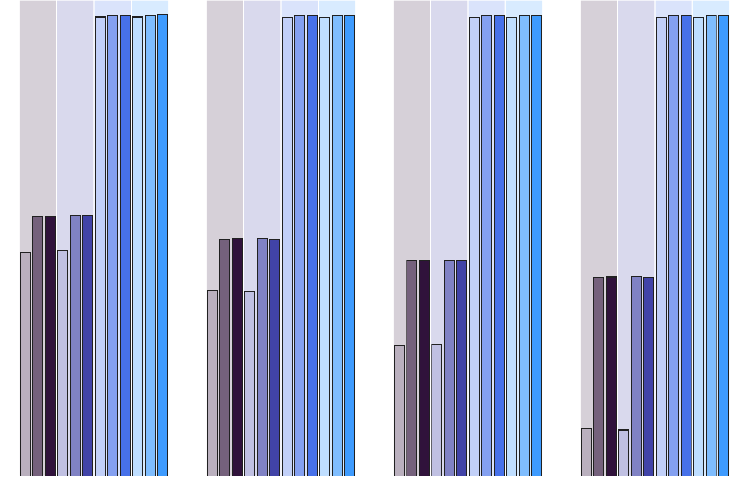};

\end{axis}
\end{tikzpicture}%
}%
	\vspace{-3mm}%
	\caption{%
		Elapsed times of the AR-based methods in the declipping experiment from Sec.\ \ref{ssec:comparison.survey} and Fig.\ \ref{fig:survey.test}.
	}
	\label{fig:survey.test:time}
\end{figure}

To evaluate the complexity numerically,
Fig.\ \ref{fig:survey.test:time} complements the results of Fig.\ \ref{fig:survey.test} in terms of elapsed times.
%Nově dopsal PR:
While regularized declipping steadily requires above two minutes of computation per a~second of audio, inpainting and GLP become faster with a~decreasing number of degraded audio samples.

To compare the observations with the reference methods,
we provide several comments taken from \cite[Sec.\,V.D]{ZaviskaRajmicOzerovRencker2021:Declipping.Survey}:
The NMF method, as one of the best performing, takes on average 30 minutes (1\,800 s) per second of input audio;
on the other hand, SS\,PEW takes only around 120~s while also scoring high values of both SDR and ODG.
The fastest methods considered in \cite{ZaviskaRajmicOzerovRencker2021:Declipping.Survey} were DR and CP with 20~seconds of computation per one second of input audio.

%\todo{PR: Slabina je, že to tehdy mohlo být počítáno na úplně jiném HW (NMF navíc počítal Alexey) \\ OM: Vím o tom, alibisticky bych to nechal a počkal, co na to recenze.}
%
Further comments and analysis are provided in Appendix~\ref{sec:elapsed.times}.%

\subsection{Software, data \& reproducible research}
\label{ssec:software}

Our software %is available in the repository
\edt{and supplementary materials (figures, listening examples) are available through the link}
\url{https://github.com/ondrejmokry/RegularizedAutoregression}.
We run the codes using Matlab R2025b.
%\todo{OM: Už i R2025b. V těch dřívějších např.\ není STOI.\\ PR: Navrhuju napsat rovnou jenom 2025b.}
%\todo{\textbf{PR:} Ověřit}
%\todo{\textbf{PR:} Napsat sem konfiguraci počítače.}

To obtain results via the basic Janssen method
%\todo[disable]{\textbf{PR:} používáme zde gapwise nebo framewise? citujeme se? \\ \textbf{OM:} framewise, na declipipng nám gapwise nebude fungovat}%
(i.e., effectively audio inpainting) we used parts of the SMALLbox~\cite{Adler2012:Audio.inpainting}.
The same applies to the GLP method 
\cite{AtlasClark2012:Generalized.linear.prediction},
whose fundamental step coincides with that of the Janssen method.
The FFT-based acceleration of the Douglas--Rachford algorithm
%DRA
was taken from \cite{Bayram2015:Structured}.
When comparing our results with those of the reference methods from the survey \cite{ZaviskaRajmicOzerovRencker2021:Declipping.Survey},
we took directly the reported and publicly available signals from the accompanying repository \url{https://github.com/rajmic/declipping2020_codes}.
%Finally, in the case of the methods based on the \qm{replace reliable} strategy, 
%we downloaded them from the webpage \url{https://rajmic.github.io/declipping2020/}.
%\todo{PR: smazat větu, pokud se rozhodneme replace nezahrnovat}

%\ond{
%\begin{itemize}
%	\item všechno bude na GitHubu
%	\item ocitujeme použití cizích kódů
%	\begin{itemize}
%		\item Janssen v případě inpaintingu a GLP podle \cite{Adler2012:Audio.inpainting},
%		\item FFT-accelerated DRA taken from \cite{Bayram2015:Structured},
%		\item results of the reference methods are taken directly from the survey \cite{ZaviskaRajmicOzerovRencker2021:Declipping.Survey} and the accompanying repository \url{https://github.com/rajmic/declipping2020_codes}, and from the webpage \url{https://rajmic.github.io/declipping2020/} in the case of the \qm{replace reliable} strategy
%	\end{itemize}
%	\item verze Matlabu (původní data R2021a, nová data R2023a), konfigurace počítače?
%\end{itemize}}

\section{Conclusion}
\label{sec:conclusion}

%In audio and speech processing,
%several approaches exist that enhance the basic popular AR model,
%either constraining the signal values in the time domain or altering the AR coefficients.
%While these approaches offer substantial benefits,

We have developed a~comprehensive and general framework
alongside its associated optimization problem and algorithm
which encompasses the previous attempts
to enhance the basic popular AR model.
The framework allows constraining the signal values in the time domain and/or regularizing the AR coefficients.
%We thoroughly discussed the computational implications of this algorithm and assessed the impact of several enhancements on its performance in terms of computational load.
%\todo{PR: Výpočetní náročnost se řeší ob jeden odstavec, tady bych to ještě nedával}
\edt{One of the novelties of our approach lies in the fact that several previously published restoration formulations become special cases of our framework.} %PR
%\todo{Možná napsat ještě něco jako že to mj. umožňuje aby starší přístupy byly ještě vylepšeny protože teďka můžou regularizovat ve více oblastech zároveň?}
\edt{As a~consequence, such methods can be reconsidered by augmenting them with an additional regularization.} %PR

The experiments focused on the problem of audio declipping as a~natural use case of the proposed framework,
where degraded signal samples are constrained to exceed a~known clipping level.
We have demonstrated the applicability of the regularized AR method and conducted a~comparison with current state-of-the-art methods.
% (including the GLP algorithm).
\edt{We have proven the competitiveness of the proposed solution in declipping of music,
%particularly in the context of signals that have experienced mild clipping.
and we have shown its superiority in declipping on a~speech dataset.}
%\todo{PR: Mělo by se aktualizovat, že na řeči si vedeme ještě lépe než na hudbě a že taky ukazujeme dekvantizaci jako příklad flexinility.}

We have also pointed out the main drawback of the regularized AR modeling,
which is the computational complexity.
In the presented declipping application,
this was mainly due to the choice of rather large dimensionality of both the coefficient and signal domains of the central problem \eqref{eq:AAR.declipping}.
While we have proposed several strategies allowing a~substantial reduction of the computational times,
further acceleration for audio processing applications
remains a~challenge.

Regarding other future directions,
the relationship of the AR coefficients and the signal,
%$\a$ and $\x$,
invoked by the minimization~\eqref{eq:AR}, could be relaxed such that the noise (error)
% $\e$
is shaped according to the masking properties of the human ear \cite[Sec.\,4.1]{AudioSignalProcessingAndCoding}. % which remains for the future work.
Such modification could lead to a~higher perceptual quality of the solution when using the proposed AR framework in the reconstruction of degraded audio signals.

% or


\appendices

\section{Acceleration}
\label{sec:acceleration}
%\todo{PR: Přílohy až za literaturou? \\ OM: Dle kontroly posledních příspěvků v TASLP appendixy buď nejsou, nebo jsou před literaturou.}

This appendix presents different strategies to accelerate the solution to the problem \eqref{eq:AAR.declipping}, following section \ref{sec:algorithms.for.subproblems}. %\ref{sec:algorithm}.
Two different approaches are presented in sections \ref{ssec:progressive.iteration} and \ref{ssec:extrapolation} and evaluated numerically in section \ref{ssec:acceleration}.
%In particular, section \ref{ssec:progressive.iteration} proposes to gradually change iterations of the solver of the subproblems \eqref{eq:ACS2:a} and \eqref{eq:ACS2:x}.

%\todo[inline]{\textbf{PR:} Nedávají se přílohy až za literatúru? V téhle šabloně ne, ale je to šablona pro TASLP?}
% OM: máme to dobře podle https://journals.ieeeauthorcenter.ieee.org/wp-content/uploads/sites/7/IEEE-Editorial-Style-Manual-for-Authors.pdf, str. 12

%OM: zmíněno už na začátku, takže bych to neopakoval
%
%\subsection{FFT-accelerated DRA}
%
%\textbf{akcelerace DRA} podle \cite{Bayram2015:Structured}

\subsection{Progressive sub-iteration counts}
\label{ssec:progressive.iteration}

Inspired by the trade-off behavior between outer and inner iteration counts (see Sec.\ \ref{ssec:iteration.tradeoff} and Fig.\ \ref{fig:tradeoff}),
we propose a~progressive strategy of choosing the inner iteration counts.
Since it appears from the experiments that high precision is necessary for a~satisfactory estimation in a single outer iteration,
the idea is to \qm{spare} some inner iterations only in the beginning of the ACS while keeping the necessary precision in the final iterations,
where the resulting signal is sought.
More specifically,
a~logarithmic distribution of the inner iterations is proposed.
Denote $I$ the number of outer iterations and $N^{(i)}$ the number of inner iterations for $i=1,\dots,I$.
For the chosen values of $n^{(1)}$ and $n^{(I)}$\!,
the proposed scheme is
\begin{equation}
	N^{(i)} = 10^{n^{(1)} + \frac{i-1}{I-1} (n^{(I)}-n^{(1)})},
\end{equation}
or, in Matlab notation%
\footnote{\url{https://www.mathworks.com/help/matlab/ref/logspace.html}}\!,
\begin{equation}
	[N^{(1)}\!, \dots, N^{(I)}] = \texttt{logspace}(n^{(1)}\!, n^{(I)}\!, I).
\end{equation}
In particular,
it holds $N^{(1)} = 10^{n^{(1)}}$ and $N^{(I)}=10^{n^{(I)}}$\!.
%\todo{PR: Zde bych čekal nějaký výsledek a ona začne další podsekce...}

\subsection{Extrapolation and line search in the sub-problems}
\label{ssec:extrapolation}

Although the acceleration of DRA helps to significantly reduce the computational time,
especially in the signal-estimation sub-problem,
% (see Fig.\ \ref{fig:gradient.versus.proximal:signal}),
the whole ACS algorithm remains very demanding.
%Replacing DRA with PGD
%\todo{\textbf{OM:} Nezavedená zkratka.}%
%\todo{\textbf{OM:} Nechceme zmínku o PGD úplně vyhodit?\\ *: vyhodit}%
%while increasing the number of iterations of the subsolver represents a possibility.
As outlined in Sec.\ \ref{ssec:iteration.tradeoff}, the reason is
that an accurate solution of the sub-problems is necessary
to achieve satisfactory results of the ACS algorithm.
%For this reason, no significant reduction of computational time is possible by any of the considered variants.
%as will be demonstrated in Sec.\ \ref{ssec:iteration.tradeoff}, the cruciality of the numerical precision does not allow for a significant computational time reduction.
%\todo{\textbf{PR:} poslední část je mi nejasná (cruciality of...) \\ OM: zkusil jsem vylepšit}

In response to this observation, we propose several extrapolation strategies for the sub-problems, as well as for the whole ACS:
\begin{enumerate}
	\item \textbf{extrapolate the coefficient update}:
	\begin{itemize}
		\item compute $\a^{(i-\frac{1}{2})}$ as a solution to
		%the RHS
		\eqref{eq:ACS2:a},
		\item extrapolate $\a^{(i)} = (1+\tcoef)\a^{(i-\frac{1}{2})} - \tcoef \a^{(i-1)}$
		\\ for the chosen step size $\tcoef > 0$,
	\end{itemize}
	\item \textbf{extrapolate the signal update}:
	\begin{itemize}
		\item compute $\x^{(i-\frac{1}{2})}$ as a solution to
		%the RHS
		\eqref{eq:ACS2:x},
		\item extrapolate $\x^{(i)} = (1+\tsig)\x^{(i-\frac{1}{2})} - \tsig \x^{(i-1)}$
		\\ for the chosen step size $\tsig > 0$,
	\end{itemize}
	\item \textbf{line search}:
	\begin{itemize}
		\item compute $\a^{(i-\frac{1}{2})}$ as a solution to
		%the RHS
		\eqref{eq:ACS2:a},
		\item compute $\x^{(i-\frac{1}{2})}$ as a solution to
		%the RHS
		\eqref{eq:ACS2:x} with $\a^{(i-\frac{1}{2})}$ fixed instead of $\a^{(i)}$\!,
		\item choose $\hat{\tau}$ such that
		\begin{equation}
			\hat{\tau} = \argmin_{\tau \geq 0}
%			\left\{ %Q_{\tau}(\tau) = 
			\,Q(\a(\tau),\x(\tau)),
%			\right\},
		\end{equation}
		\todo[disable]{\textbf{PR:} Má smysl to divné dvojité tau vůbec zavádět? \\ \textbf{OM:} asi by šlo jenom popsat\\ PR: Nebo vynechat to před rovnítkem}%
		where $\a(\tau) = (1+\tau)\a^{(i-\frac{1}{2})} - \tau \a^{(i-1)}$\\
		 and $\x(\tau) = (1+\tau)\x^{(i-\frac{1}{2})} - \tau \x^{(i-1)}$,
		\item extrapolate as $\a^{(i)} = \a(\hat{\tau})$ and $\x^{(i)} = \x(\hat{\tau})$.
	\end{itemize}
\end{enumerate}

A problem of the line search strategy is the non-convexity%
\footnote{%
	In the context of the line search, an important observation is that the objective is non-convex over the half-line under consideration.
	The only exception is a discrete set of search directions, as can be observed in an illustrative example
	of $Q(a,x) = a\cdot x$ with the scalars $a,x$.
}
of the composed problem \eqref{eq:AAR}.
Thus, no guarantee of finding a~minimum with respect to the considered half-line is at hand in general.
However, our empirical examples show that an improvement of the objective via sampling a suitable selection of the half-line is possible.
%\todo{\textbf{OM:} Upraveno z textu níže. Šel by přidat i obrázek, ale nechci to zahltit.}
\todo[inline,disable]{\textbf{PR:} Ok, to je čistě Tvůj empirický postřeh nebo na to existuje nějaká teorie? \\
	\textbf{OM:} záruka podle mě není, právě kvůli nekonvexitě; empiricky to mám pěkně ověřené na tom minipříkladku $Q(a,x) = a\cdot x$ (skript \texttt{linesearch\_meaningfulness.m}),
	případně jde pustit \texttt{demo.m} s nastavením \texttt{linesearch = true} a \texttt{plotLS = true}}%

%\todelete{
	%\begin{itemize}
	%	\item
	%		ilustračním příkladem na modelové funkci
	%		\begin{equation}
		%			Q(a,x) = a\cdot x
		%		\end{equation}
	%		lze nahlédnout, že v případě line search je účelová funkce podél směru hledání nekonvexní
	%		(konkrétně $Q(a,x)$ je konvexní pouze podél přímek procházejících středem nebo pro $a$ nebo $x$ konstantní)
	%		\todo{\textbf{PR:} Ok, může se zmínit, ale spíš jen stručně, ať si nad tím každý pošpekuluje sám :)}
	%	\item
	%		v praxi však lze dosáhnout zlepšení hodnoty účelové funkce, protože překonáme interval nekonvexity a v některých případech za ním najdeme globální minimum (globální vzhledem k polopřímce)
	%\end{itemize}}

\subsection{Numerical experiment}
\label{ssec:acceleration}

\begin{figure*}
	\centering
	\adjustbox{width=\linewidth}{\input{figures/accelerations.tex}}
	\caption{%
		Comparison of the acceleration options for the frame length $w=8192$ samples and the AR model order $p=512$.
		The smaller and lighter points represent individual reconstructions in the experiment, the larger points are the mean values.
	}
	\label{fig:accelerations}
\end{figure*}

The acceleration strategies proposed above have been tested on a~declipping problem with an input SDR of 10\,dB (see Sec.\ \ref{sec:experiments} for details).
As the baseline serves the ACS algorithm with $I=10$ outer iterations and DRA as the sub-solver with 1\,000 iterations.
The accelerated versions include:
\begin{itemize}
	\item FFT-based acceleration of the DRA \cite{Bayram2015:Structured}.
	\item Progressive iteration according to Sec.\ \ref{ssec:progressive.iteration} with 100 iterations of DRA in the first iteration of ACS ($n^{(1)}=2$, $N^{(1)}=100$) and 1\,000 iterations of DRA in the last iteration of ACS ($n^{(10)}=3$, $N^{(10)}=1\,000$), spaced logarithmically.
	\item Extrapolation according to Sec.\ \ref{ssec:extrapolation} with the following extrapolation steps for the signal and/or the AR model coefficients in the iteration $i$ of $I=5$:
	\begin{subequations}
		\label{eq:extra}
		\begin{align}
			\x^{(i)} &= \x^{(i-\frac{1}{2})} + \tfrac{I-i}{I-1}(\x^{(i-\frac{1}{2})} - \x^{(i-1)}),
			\label{eq:extra:signal}
			\\
			\a^{(i)} &= \a^{(i-\frac{1}{2})} + 2\tfrac{I-i}{I-1}(\a^{(i-\frac{1}{2})} - \a^{(i-1)}).
			\label{eq:extra:coefs}
		\end{align}
	\end{subequations}
	In line with \ref{ssec:extrapolation}, we have in the $i$-th iteration $\tsig = \tfrac{I-i}{I-1}$ (i.e., the extrapolation step length decreases from $\tsig=1$ in the first iteration to $\tsig=0$, meaning no extrapolation in the $I$-th iteration).
	Similarly, we have $\tcoef = 2\tfrac{I-i}{I-1}$.
	\item Line search according to Sec.\ \ref{ssec:extrapolation}, considering only $\tau\in[0.0001, 100]$ for computational reasons.
\end{itemize}
To evaluate the acceleration,
the computation time is tracked for the considered ways of estimating the AR coefficients and the declipped signal, together with the reconstruction quality.
The experiment was performed on a~PC with the Intel Core~i7 3.40\,GHz processor, 32\,GB RAM.

The numerical results for five short signals\footnote{%
	The test signals are single-instrument recordings extracted from the set based on the EBU SQAM database, as described in Sec.\ \ref{sec:experiments},
	and shortened to a~length of 0.6\,s.
}
% a~subset of five test signals
%\todo{[OM] pozor, je jich nejen míň, ale ještě jsou zkrácené}
%\todo{PR: Vyjasnit jestli je to subset [OM]}
are presented in Fig.\ \ref{fig:accelerations} and \ref{fig:accelerations.bars}.
Fig.\ \ref{fig:accelerations} shows the individual reconstructions in the time-quality plane, where the reconstruction quality is measured using SDR.
%\todo{\textbf{PR:} Zvýšil bych tučnost všech křížků a koleček \\ \textbf{OM:} co teď?}
In this plot, a suitable acceleration strategy should have its data points to the left from the baseline (meaning lower computational time) and not below the baseline (meaning the same or a~better reconstruction quality).
The same data are
%\todo[color=orange!25!white]{PR: singulár či plurál? [dr. Kříž]}
% OM: vždy plurál podle https://journals.ieeeauthorcenter.ieee.org/wp-content/uploads/sites/7/IEEE-Editorial-Style-Manual-for-Authors.pdf, str. 22
% PR: "The word “data” is plural, not singular."...citace ze sekce "Some Common Mistakes" šablony pro ICASSP25
shown in Fig.~\ref{fig:accelerations.bars} in a~different manner.
There, the decrease in computational time (i.e., time gain) and decrease in SDR (i.e., performance loss) are shown together as stacked bars for each acceleration strategy.
Furthermore, this plot shows the data for different orders of the AR model and different frame lengths, since the acceleration rate is supposed to depend on the problem size.
Note that all the values shown in Fig.\ \ref{fig:accelerations.bars} are relative with respect to the baseline values shown in Fig.\ \ref{fig:accelerations.bars.baseline}.

The conclusion from this experiment is that the most beneficial option is to use the FFT-accelerated DRA in both sub-problems of \eqref{eq:ACS2}.
%i.e., the (c) variant in inpainting (and GLP, since these do not differ significantly in the computational load) and the (c, s) variant in declipping.
The computation time is greater merely in some cases of inpainting and GLP.
%, which is also in line with the computational times of the AR model estimation presented in Fig.\ \ref{fig:gradient.versus.proximal:coefficients}.
Regarding the reconstruction quality, the PEAQ ODG values do not show any difference, and the related graphical representation is thus omitted.
In terms of PEAQ ODG or SDR, a~loss of reconstruction quality is observed;
mainly in cases including the extrapolation of the estimated AR model.
%(\ond{cases (4) and (6)}).
A~less significant quality loss is observed in the line search case
%\ond{(3)}
and signal extrapolation.
%\ond{(5)}.

As a~compromise between acceleration and reconstruction quality loss,
the experiment favors using 1\,000 inner iterations of the FFT-accelerated DRA with extrapolation of the signal update according to \eqref{eq:extra:signal}
%(case \ond{(5)} in
(see Fig.\ \ref{fig:accelerations} and \ref{fig:accelerations.bars}).

\begin{figure*}
	\centering
	\adjustbox{width=0.95\linewidth}{% This file was created by matlab2tikz.
\begin{tikzpicture}

\begin{axis}[%
width=4.521in,
height=2.0in,
at={(0.758in,0.909in)},
scale only axis,
bar shift auto,
log origin=infty,
xmin=0.509090909090909,
xmax=6.49090909090909,
xtick={1,2,3,4,5,6},
xticklabels={{$p = 512,\;w = 2048$},{$p = 512,\;w = 4096$},{$p = 512,\;w = 8192$},{$p = 2048,\;w = 2048$},{$p = 2048,\;w = 4096$},{$p = 2048,\;w = 8192$}},
xticklabel style={rotate=30},
ymode=log,
ymin=12.33093878,
ymax=10000,
yminorticks=true,
ylabel={time (s)},
axis background/.style={fill=white},
title={elapsed times of the baseline algorithms},
]
\addplot[ybar, bar width=0.145, fill=white!20!black, draw=black, area legend] table[row sep=crcr] {%
1	18.48347032\\
2	12.33093878\\
3	15.4089691\\
4	530.83728898\\
5	286.8271733\\
6	185.41669436\\
};
%\addlegendentry{inpainting}

\addplot[ybar, bar width=0.145, fill=white!40!black, draw=black, area legend] table[row sep=crcr] {%
1	532.80924104\\
2	1107.62089266\\
3	2783.84883056\\
4	1046.45905162\\
5	1382.55427506\\
6	2973.34391248\\
};
%\addlegendentry{consistent declipping}

\addplot[ybar, bar width=0.145, fill=white!60!black, draw=black, area legend] table[row sep=crcr] {%
1	529.69414244\\
2	1111.29963216\\
3	2769.85326866\\
4	1045.43059234\\
5	1379.6392508\\
6	2992.63957658\\
};
%\addlegendentry{inconsistent declipping}

\addplot[ybar, bar width=0.145, fill=white!80!black, draw=black, area legend] table[row sep=crcr] {%
1	18.24481646\\
2	12.44835624\\
3	15.32955558\\
4	530.8180748\\
5	287.12142412\\
6	185.03721166\\
};
%\addlegendentry{GLP}

\end{axis}

\begin{axis}[%
width=4.521in,
height=2.0in,
at={(6.021in,0.909in)},
scale only axis,
bar shift auto,
xmin=0.509090909090909,
xmax=6.49090909090909,
xtick={1,2,3,4,5,6},
xticklabels={{$p = 512,\;w = 2048$},{$p = 512,\;w = 4096$},{$p = 512,\;w = 8192$},{$p = 2048,\;w = 2048$},{$p = 2048,\;w = 4096$},{$p = 2048,\;w = 8192$}},
xticklabel style={rotate=30},
ymin=0,
ymax=30,
ylabel={SDR (dB)},
axis background/.style={fill=white},
title={performance of the baseline algorithms},
legend style={at={(0.97, 0.03)}, anchor=south east, legend cell align=left, align=left, draw=white!15!black}
]
\addplot[ybar, bar width=0.145, fill=white!20!black, draw=black, area legend] table[row sep=crcr] {%
1	24.7968580178273\\
2	26.912601526654\\
3	27.6872887491611\\
4	24.7603141448659\\
5	27.2466253033831\\
6	28.7095234684838\\
};
\addplot[forget plot, color=white!15!black] table[row sep=crcr] {%
0.509090909090909	0\\
6.49090909090909	0\\
};
\addlegendentry{inpainting}

\addplot[ybar, bar width=0.145, fill=white!40!black, draw=black, area legend] table[row sep=crcr] {%
1	25.4397472297678\\
2	27.4683745860625\\
3	28.4707949919423\\
4	25.1859354556597\\
5	27.8852102865643\\
6	29.5452485366732\\
};
\addplot[forget plot, color=white!15!black] table[row sep=crcr] {%
0.509090909090909	0\\
6.49090909090909	0\\
};
\addlegendentry{consistent declipping}

\addplot[ybar, bar width=0.145, fill=white!60!black, draw=black, area legend] table[row sep=crcr] {%
1	25.3274262947257\\
2	27.3011881986502\\
3	28.264205891712\\
4	25.1868254674405\\
5	27.6994671568778\\
6	29.3571189281579\\
};
\addplot[forget plot, color=white!15!black] table[row sep=crcr] {%
0.509090909090909	0\\
6.49090909090909	0\\
};
\addlegendentry{inconsistent declipping}

\addplot[ybar, bar width=0.145, fill=white!80!black, draw=black, area legend] table[row sep=crcr] {%
1	24.2188764951528\\
2	26.5940768241352\\
3	27.5316755070661\\
4	24.2524739120586\\
5	27.0258220933314\\
6	28.6235126209508\\
};
\addplot[forget plot, color=white!15!black] table[row sep=crcr] {%
0.509090909090909	0\\
6.49090909090909	0\\
};
\addlegendentry{GLP}

\end{axis}
\end{tikzpicture}%
}
	\caption{Baseline elapsed times and performance in terms of SDR using 10 ACS iterations and 1000 DRA iterations with no acceleration strategy.}
	\label{fig:accelerations.bars.baseline}
\end{figure*}
\begin{figure*}
	\centering
	\adjustbox{width=\linewidth}{% This file was created by matlab2tikz.
\definecolor{mycolor1}{rgb}{0.00000,0.44700,0.74100}%
\definecolor{mycolor2}{rgb}{0.85000,0.32500,0.09800}%
\definecolor{mycolor3}{rgb}{0.92900,0.69400,0.12500}%
\definecolor{mycolor4}{rgb}{0.49400,0.18400,0.55600}%
\definecolor{mycolor5}{rgb}{0.46600,0.67400,0.18800}%
\definecolor{mycolor6}{rgb}{0.30100,0.74500,0.93300}%
\begin{tikzpicture}

\begin{axis}[%
width=3.0in,
height=2.0in,
at={(1.389in,6.25in)},
scale only axis,
bar width=0.125,
xmin=0.4,
xmax=6.6,
xtick={1,2,3,4,5,6},
xticklabels={{$p = 512,\;w = 2048$},{$p = 512,\;w = 4096$},{$p = 512,\;w = 8192$},{$p = 2048,\;w = 2048$},{$p = 2048,\;w = 4096$},{$p = 2048,\;w = 8192$}},
xticklabel style={rotate=30},
ymin=-0.6,
ymax=1.2,
axis background/.style={fill=white},
title={inpainting (none)},
axis on top
]

\addplot [forget plot] graphics [xmin=-0.2375, xmax=7.2375, ymin=-1, ymax=1.13100704812175] {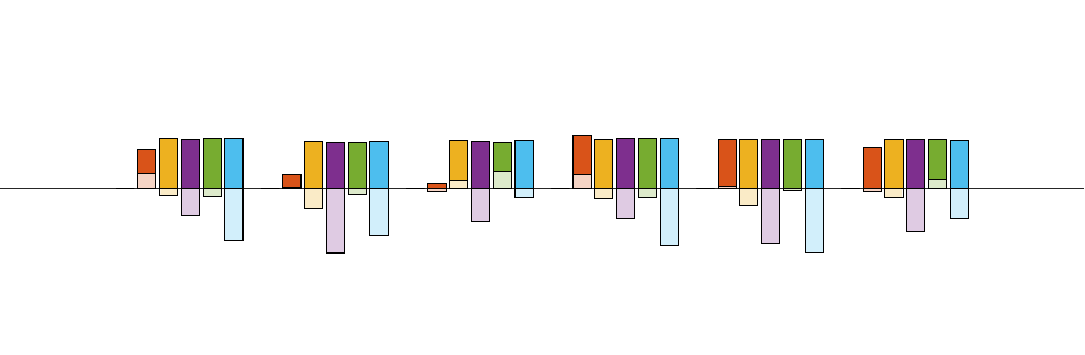};

\end{axis}

\begin{axis}[%
width=3.0in,
height=2.0in,
at={(1.389in,3.0in)},
scale only axis,
bar width=0.125,
xmin=0.4,
xmax=6.6,
xtick={1,2,3,4,5,6},
xticklabels={{$p = 512,\;w = 2048$},{$p = 512,\;w = 4096$},{$p = 512,\;w = 8192$},{$p = 2048,\;w = 2048$},{$p = 2048,\;w = 4096$},{$p = 2048,\;w = 8192$}},
xticklabel style={rotate=30},
ymin=-0.6,
ymax=1.2,
axis background/.style={fill=white},
title={inpainting (coef. sub-problem)},
axis on top
]

\addplot [forget plot] graphics [xmin=-0.2375, xmax=7.2375, ymin=-1, ymax=1.13100704812175] {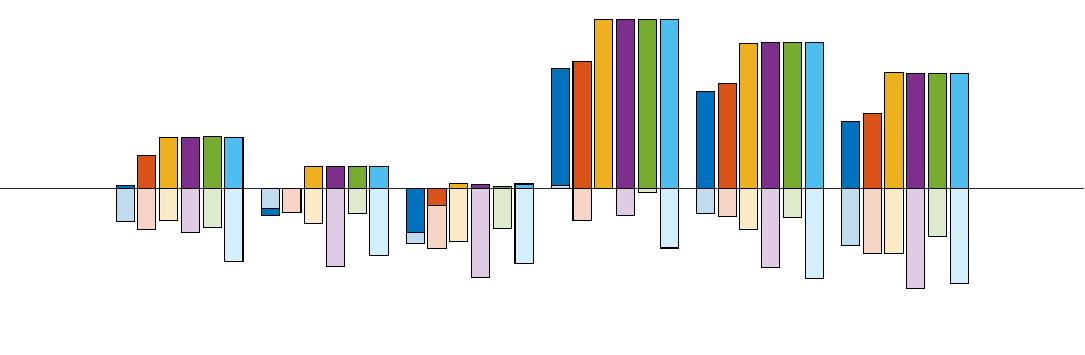};

\end{axis}

\begin{axis}[%
width=3.0in,
height=2.0in,
at={(4.889in,6.25in)},
scale only axis,
bar width=0.125,
xmin=0.4,
xmax=6.6,
xtick={1,2,3,4,5,6},
xticklabels={{$p = 512,\;w = 2048$},{$p = 512,\;w = 4096$},{$p = 512,\;w = 8192$},{$p = 2048,\;w = 2048$},{$p = 2048,\;w = 4096$},{$p = 2048,\;w = 8192$}},
xticklabel style={rotate=30},
ymin=-0.6,
ymax=1.2,
axis background/.style={fill=white},
title={consistent declipping (none)},
axis on top
]

\addplot [forget plot] graphics [xmin=-0.2375, xmax=7.2375, ymin=-1, ymax=1.13100704812175] {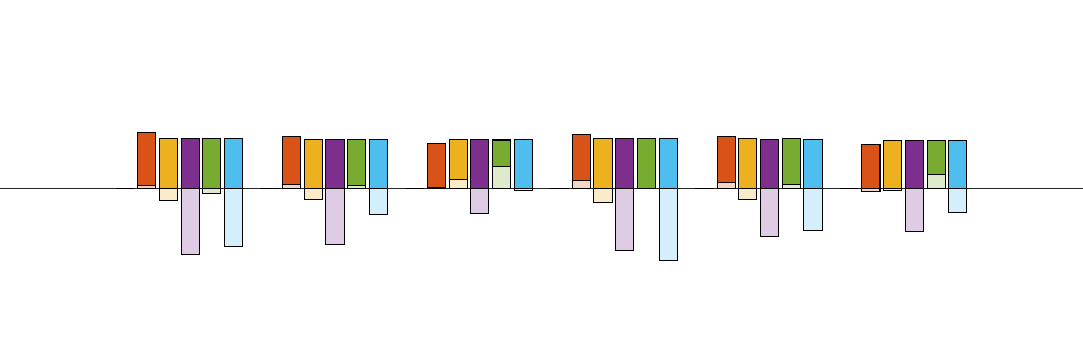};

\end{axis}

\begin{axis}[%
width=3.0in,
height=2.0in,
at={(4.889in,3.0in)},
scale only axis,
bar width=0.125,
xmin=0.4,
xmax=6.6,
xtick={1,2,3,4,5,6},
xticklabels={{$p = 512,\;w = 2048$},{$p = 512,\;w = 4096$},{$p = 512,\;w = 8192$},{$p = 2048,\;w = 2048$},{$p = 2048,\;w = 4096$},{$p = 2048,\;w = 8192$}},
xticklabel style={rotate=30},
ymin=-0.6,
ymax=1.2,
axis background/.style={fill=white},
title={consistent declipping (coef. sub-problem)},
axis on top
]

\addplot [forget plot] graphics [xmin=-0.2375, xmax=7.2375, ymin=-1, ymax=1.13100704812175] {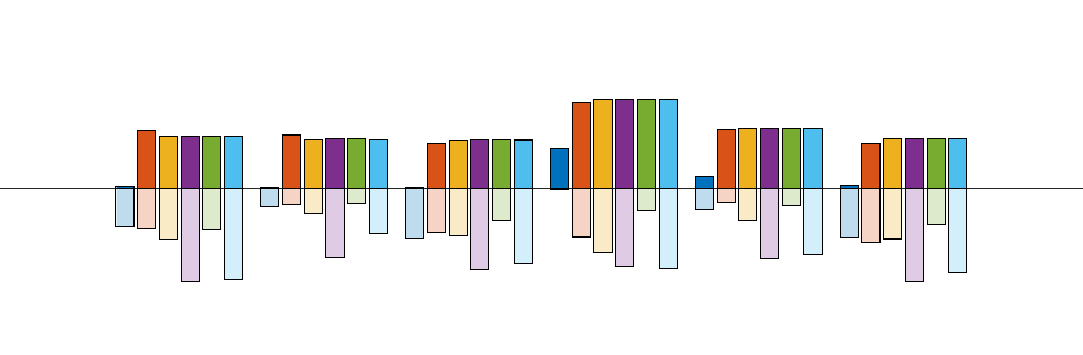};

\end{axis}

\begin{axis}[%
width=3.0in,
height=2.0in,
at={(8.389in,6.25in)},
scale only axis,
bar width=0.125,
xmin=0.4,
xmax=6.6,
xtick={1,2,3,4,5,6},
xticklabels={{$p = 512,\;w = 2048$},{$p = 512,\;w = 4096$},{$p = 512,\;w = 8192$},{$p = 2048,\;w = 2048$},{$p = 2048,\;w = 4096$},{$p = 2048,\;w = 8192$}},
xticklabel style={rotate=30},
ymin=-0.6,
ymax=1.2,
axis background/.style={fill=white},
title={consistent declipping (sig. sub-problem)},
axis on top
]

\addplot [forget plot] graphics [xmin=-0.2375, xmax=7.2375, ymin=-1, ymax=1.13100704812175] {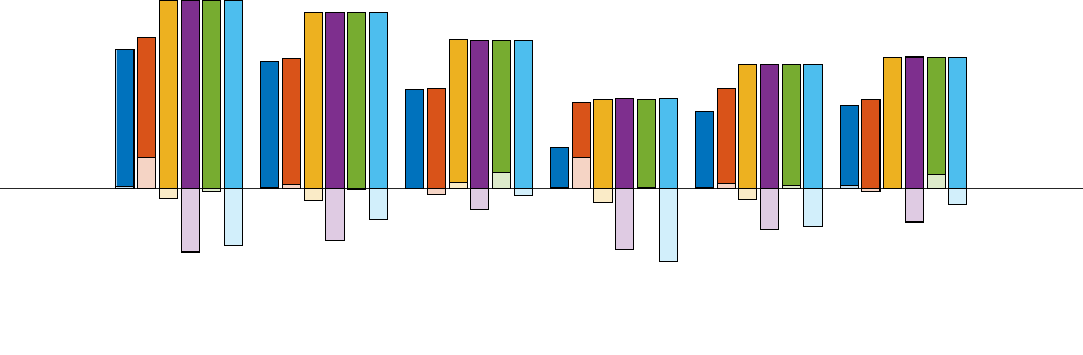};

\end{axis}

\begin{axis}[%
width=3.0in,
height=2.0in,
at={(8.389in,3.0in)},
scale only axis,
bar width=0.125,
xmin=0.4,
xmax=6.6,
xtick={1,2,3,4,5,6},
xticklabels={{$p = 512,\;w = 2048$},{$p = 512,\;w = 4096$},{$p = 512,\;w = 8192$},{$p = 2048,\;w = 2048$},{$p = 2048,\;w = 4096$},{$p = 2048,\;w = 8192$}},
xticklabel style={rotate=30},
ymin=-0.6,
ymax=1.2,
axis background/.style={fill=white},
title={consistent declipping (coef. and sig. sub-problems)},
axis on top
]

\addplot [forget plot] graphics [xmin=-0.2375, xmax=7.2375, ymin=-1, ymax=1.13100704812175] {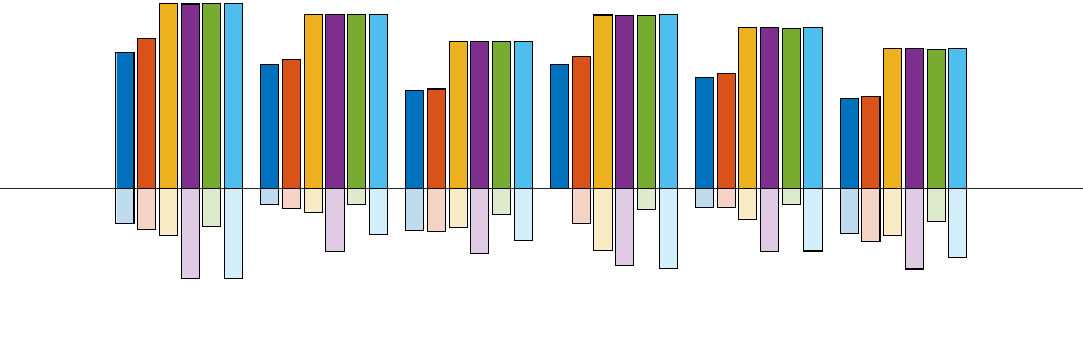};

\end{axis}

\begin{axis}[%
width=3.0in,
height=2.0in,
at={(2.0in,0.0in)},
scale only axis,
xmin=0,
xmax=1,
ymin=-1,
ymax=1,
axis line style={draw=none},
ticks=none,
axis x line*=bottom,
axis y line*=left,
legend style={at={(1.55, 0.5)}, anchor=west, legend cell align=left, align=left}
]
\addplot [color=black, forget plot]
  table[row sep=crcr]{%
0	0\\
1	0\\
};

\node[right, align=left, inner sep=0]
at (axis cs:0.05,0.6) {$\uparrow$ time gain (darker shade)};
\node[right, align=left, inner sep=0]
at (axis cs:0.08,0.35) {log10 ((baseline time) / (accelerated time))};
\node[right, align=left, inner sep=0]
at (axis cs:0.08,0.15) {positive value = acceleration works};
\node[right, align=left, inner sep=0]
at (axis cs:0.08,-0.15) {negative value = performance decreases};
\node[right, align=left, inner sep=0]
at (axis cs:0.08,-0.35) {(accelerated SDR (dB)) -- (baseline SDR (dB))};
\node[right, align=left, inner sep=0]
at (axis cs:0.05,-0.6) {$\downarrow$ performance loss (lighter shade)};

\addlegendimage{empty legend}
\addlegendentry{\hspace{-2em}\textbf{acceleration strategies}}

\addlegendimage{draw=black, fill=mycolor1, area legend}
\addlegendentry{\hspace{0.5em}10 ACS iterations, 1000 DRA iterations}

\addlegendimage{draw=black, fill=mycolor2, area legend}
\addlegendentry{\hspace{0.5em}10 ACS iterations, progressive DRA iterations}

\addlegendimage{draw=black, fill=mycolor3, area legend}
\addlegendentry{\hspace{0.5em}5 ACS iterations, 1000 DRA iterations, line search}

\addlegendimage{draw=black, fill=mycolor4, area legend}
\addlegendentry{\hspace{0.5em}5 ACS iterations, 1000 DRA iterations, coef. extrapolation}

\addlegendimage{draw=black, fill=mycolor5, area legend}
\addlegendentry{\hspace{0.5em}5 ACS iterations, 1000 DRA iterations, sig. extrapolation}

\addlegendimage{draw=black, fill=mycolor6, area legend}
\addlegendentry{\hspace{0.5em}5 ACS iterations, 1000 DRA iterations, coef. and sig. extrapolation}

\end{axis}

\end{tikzpicture}%
}
	\caption{%
		Performance of the acceleration strategies for an input SDR of 10\,dB, 
		dependent on the frame length $w$ and the AR model order $p$.
		Color coding is the same in the whole figure and is identical to Fig.\ \ref{fig:accelerations}.
		FFT-acceleration of the DRA in the coefficient and signal sub-problems \eqref{eq:ACS2:a} and \eqref{eq:ACS2:x} is indicated in brackets above the individual plots.
		Positive time gain (darker-shade bar has a~value above 0) means that the acceleration truly reduces computation time.
		A~positive performance loss (the lighter-shade bar has a~value below 0) means that the acceleration causes a~drop in the reconstruction SDR. %of the reconstructed signal.
		For the baseline time and SDR, see Fig.\ \ref{fig:accelerations.bars.baseline}.
	}
	\label{fig:accelerations.bars}
\end{figure*}

\section{Declipped signal consistency}
\label{sec:consistency}

While the proposed signal regularization rationalizes the use of AR model for audio declipping,
the results do not show large improvement over the inpainting or GLP techniques in many cases.
A hypothesis is that
although consistency in the time domain (satisfying the clipping conditions) can be approached with $\lsig>0$ and even theoretically guaranteed in the case of $\lsig = \infty$,
solving the sub-problem \eqref{eq:ACS2:x} in an iterative manner carries the risk of reaching only an approximate solution.
Furthermore,
consistency does not necessarily ensure higher SDR.

To support these ideas,
we analyze the resulting signals from the experiment presented in Sec.\ \ref{ssec:comparison.survey} and relate the $\Delta$SDR scores (see also Eq.\ \eqref{eq:sdr}) to the consistency of the signals.
For any solution $\hat{\x}$, we measure the (squared) distance from the consistency set
$\frac{1}{2}d_{\setdec}(\hat{\x})^2 = \frac{1}{2}\norm{\hat{\x}-\proj_{\setdec}(\hat{\x})}^2$,
which is also part of the objective of the problem \eqref{eq:AAR.declipping} in the declipping case.
The assorted results for all audio samples and all values of input SDR are shown in Fig.\ \ref{fig:consistency}.

\begin{figure}[ht]
	\centering
	\adjustbox{scale=0.5}{\input{figures/consistency.tex}}%
	\vspace{-2mm}%
	\caption{%
		Consistency of the resulting signals from the experiment from Sec.~\ref{ssec:comparison.survey} and Fig.~\ref{fig:survey.test}.
		Note that the values are cropped at the minimal value of $10^{-12}$, considered to be essentially zero.
	}
	\label{fig:consistency}
\end{figure}

Importantly,
GLP reaches similar values of $d_{\setdec}(\hat{\x})^2$ compared to inpainting,
which means that in our experiments, the heuristic signal rectification step in GLP does not guarantee consistency of the result.
On the other hand,
the declipping strategy provides solutions that tend towards lower values of $d_{\setdec}(\hat{\x})^2$, approaching even (numerical) zero in the case of $\lsig=\infty$.

Regarding the relation of consistency and reconstruction quality,
Fig.\ \ref{fig:consistency} reveals some correlation in the case of inpainting and GLP.
This is natural in the sense that very high SDR values correspond to the solution $\hat{\x}$ being close to the ground truth signal, which is necessarily consistent.
In the declipping case, (approximate) consistency was ensured by definition of the problem, thus no further correlation with $\Delta$SDR is observed.

\section{Elapsed times in a demonstrative example}
\label{sec:elapsed.times}

Fig.\ \ref{fig:demo.times} shows an illustrative example on the computation times, depending on several parameters.
The reconstruction task concerns a single realization of an AR process of length $N$ samples which was peak-normalized and clipped depending on the threshold $\tc$.
Different variants of the ACS algorithm are employed (using 1\,000 inner iterations of the FFT-accelerated DRA), as in Sec.\ \ref{sec:experiments}, and elapsed time is measured.

The plots illustrate in more detail several expected facts, following up on Sec.\ \ref{ssec:complexity} and also extending the results in Fig.\ \ref{fig:survey.test:time} and \ref{fig:accelerations.bars.baseline}:
\begin{itemize}
	\item The declipping task is numerically more demanding than either inpainting or GLP, independent of the value of $\lsig$.
	\item The regularization of the AR coefficients makes the estimation more demanding compared to standard methods like the Levinson--Durbin algorithm.
	\item The complexity grows with the model order (see the first plot in Fig.\ \ref{fig:demo.times}) and signal/window length (second plot in Fig.\ \ref{fig:demo.times}).
	%\todo{PR: Opět dopsat, že je to v prvním případě lineární a v druhém kvadratické nebo kubické? \\ OM: Tak jednoduché to není (viz tabulka na papíře). Otázka tudíž je, zda to chceme rozepisovat (a případně co smažeme, aby se to vešlo).}
	%
	\item The clipping threshold (or number of missing samples) only affects the complexity of the inpainting and GLP methods. The load grows with the number of missing samples, i.e., it decreases with $\tc$ (see the third plot in Fig.\ \ref{fig:demo.times}).
\end{itemize} 

\begin{figure*}
	\centering
	\adjustbox{width=\linewidth}{% This file was created by matlab2tikz.
\definecolor{mycolor1}{rgb}{0.00000,0.44700,0.74100}%
\definecolor{mycolor2}{rgb}{0.85000,0.32500,0.09800}%
\definecolor{mycolor3}{rgb}{0.92900,0.69400,0.12500}%
\definecolor{mycolor4}{rgb}{0.49400,0.18400,0.55600}%
\begin{tikzpicture}

\begin{axis}[%
width=2.000in,
height=2.000in,
at={(1.206in,0.501in)},
scale only axis,
xmin=0,
xmax=2048,
xlabel={model order $p$},
ymin=0,
ymax=1.5,
ylabel={elapsed time per outer iteration (s)},
axis background/.style={fill=white},
title style={align=center},
title={dependence on model order ($p$) \\ for fixed $N=4096$ and {\normalfont $\tc=0.2$}},
xmajorgrids,
xminorgrids,
ymajorgrids,
yminorgrids
]
\addplot [color=mycolor1, forget plot]
  table[row sep=crcr]{%
16	0.07045225\\
32	0.08713841\\
64	0.07698663\\
128	0.08613686\\
256	0.10069603\\
512	0.12310528\\
1024	0.1587277\\
2048	0.18128144\\
};
\addplot [color=mycolor1, dashed, line width=1.0pt, forget plot]
  table[row sep=crcr]{%
16	0.20895561\\
32	0.20716672\\
64	0.20492428\\
128	0.22377848\\
256	0.2969586\\
512	0.28377176\\
1024	0.35821651\\
2048	0.57907712\\
};
\addplot [color=mycolor2, forget plot]
  table[row sep=crcr]{%
16	0.06966861\\
32	0.07826479\\
64	0.07643626\\
128	0.08930966\\
256	0.09608503\\
512	0.11711093\\
1024	0.15069558\\
2048	0.17872122\\
};
\addplot [color=mycolor2, dashed, line width=1.0pt, forget plot]
  table[row sep=crcr]{%
16	0.19528346\\
32	0.2011195\\
64	0.19916155\\
128	0.21803326\\
256	0.37704505\\
512	0.27923668\\
1024	0.34783727\\
2048	0.58358624\\
};
\addplot [color=mycolor3, forget plot]
  table[row sep=crcr]{%
16	0.77869596\\
32	0.76839304\\
64	0.75590666\\
128	0.81305989\\
256	0.83885064\\
512	0.79930646\\
1024	0.8467734\\
2048	0.99868236\\
};
\addplot [color=mycolor3, dashed, line width=1.0pt, forget plot]
  table[row sep=crcr]{%
16	0.90326983\\
32	0.88297293\\
64	0.9151486\\
128	0.96198344\\
256	1.02840438\\
512	1.02386795\\
1024	1.08649047\\
2048	1.42558565\\
};
\addplot [color=mycolor4, forget plot]
  table[row sep=crcr]{%
16	0.7722683\\
32	0.78057111\\
64	0.77763808\\
128	0.80131905\\
256	0.84932615\\
512	0.84190215\\
1024	0.92562884\\
2048	1.15656655\\
};
\addplot [color=mycolor4, dashed, line width=1.0pt, forget plot]
  table[row sep=crcr]{%
16	0.95771519\\
32	0.91628253\\
64	0.91136981\\
128	0.94101225\\
256	1.03116097\\
512	0.99480095\\
1024	1.13597295\\
2048	1.4564896\\
};
\end{axis}

\begin{axis}[%
width=2.000in,
height=2.000in,
at={(3.83in,0.501in)},
scale only axis,
xmin=0,
xmax=8192,
xlabel={segment length $N$},
ymin=0,
ymax=5,
ylabel={elapsed time per outer iteration (s)},
axis background/.style={fill=white},
title style={align=center},
title={dependence on segment length ($N$) \\ for fixed $p=256$ and {\normalfont $\tc=0.2$}},
xmajorgrids,
xminorgrids,
ymajorgrids,
yminorgrids
]
\addplot [color=mycolor1, forget plot]
  table[row sep=crcr]{%
512	0.00476389\\
1024	0.0123661\\
2048	0.0261174\\
4096	0.10069603\\
8192	0.40140481\\
};
\addplot [color=mycolor1, dashed, line width=1.0pt, forget plot]
  table[row sep=crcr]{%
512	0.05177922\\
1024	0.06898616\\
2048	0.1099387\\
4096	0.2969586\\
8192	0.68068593\\
};
\addplot [color=mycolor2, forget plot]
  table[row sep=crcr]{%
512	0.00503197\\
1024	0.0122278\\
2048	0.02698778\\
4096	0.09608503\\
8192	0.41435779\\
};
\addplot [color=mycolor2, dashed, line width=1.0pt, forget plot]
  table[row sep=crcr]{%
512	0.0429211\\
1024	0.06680816\\
2048	0.10840534\\
4096	0.37704505\\
8192	0.64257421\\
};
\addplot [color=mycolor3, forget plot]
  table[row sep=crcr]{%
512	0.04891969\\
1024	0.08433348\\
2048	0.20233193\\
4096	0.83885064\\
8192	4.20718568\\
};
\addplot [color=mycolor3, dashed, line width=1.0pt, forget plot]
  table[row sep=crcr]{%
512	0.08805436\\
1024	0.13504239\\
2048	0.28242931\\
4096	1.02840438\\
8192	4.56975628\\
};
\addplot [color=mycolor4, forget plot]
  table[row sep=crcr]{%
512	0.04873009\\
1024	0.07963472\\
2048	0.19452611\\
4096	0.84932615\\
8192	4.22167045\\
};
\addplot [color=mycolor4, dashed, line width=1.0pt, forget plot]
  table[row sep=crcr]{%
512	0.08329103\\
1024	0.13762343\\
2048	0.28750866\\
4096	1.03116097\\
8192	4.47671587\\
};
\end{axis}

\begin{axis}[%
width=2.000in,
height=2.000in,
at={(6.553in,0.501in)},
scale only axis,
xmin=0.1,
xmax=0.8,
xtick distance=0.1,
xlabel={clipping threshold $\tc$},
ymin=0,
ymax=1.2,
ylabel={elapsed time per outer iteration (s)},
axis background/.style={fill=white},
title style={align=center},
title={dependence on clipping threshold ({\normalfont $\tc=0.2$}) \\ for fixed $N=4096$ and $p=256$},
xmajorgrids,
xminorgrids,
ymajorgrids,
yminorgrids,
legend style={at={(1.1,0.5)}, anchor=west, legend cell align=left, align=left}
]
\addplot [color=mycolor1]
  table[row sep=crcr]{%
0.1	0.16898813\\
0.2	0.10069603\\
0.3	0.06030136\\
0.4	0.03781343\\
0.5	0.01589533\\
0.6	0.01035738\\
0.7	0.0056406\\
0.8	0.00303263\\
};
\addlegendentry{inp., $\lcoef=0$}

\addplot [color=mycolor1, dashed, line width=1.0pt]
  table[row sep=crcr]{%
0.1	0.35917833\\
0.2	0.2969586\\
0.3	0.25649472\\
0.4	0.23378042\\
0.5	0.22507143\\
0.6	0.21017903\\
0.7	0.22000349\\
0.8	0.20244342\\
};
\addlegendentry{inp., $\lcoef=0.1$}

\addplot [color=mycolor2]
  table[row sep=crcr]{%
0.1	0.1559404\\
0.2	0.09608503\\
0.3	0.05770812\\
0.4	0.03254031\\
0.5	0.01617678\\
0.6	0.0110714\\
0.7	0.00475904\\
0.8	0.00303022\\
};
\addlegendentry{GLP, $\lcoef=0$}

\addplot [color=mycolor2, dashed, line width=1.0pt]
  table[row sep=crcr]{%
0.1	0.36023015\\
0.2	0.37704505\\
0.3	0.25135849\\
0.4	0.22518723\\
0.5	0.2144251\\
0.6	0.202028\\
0.7	0.20205976\\
0.8	0.20657111\\
};
\addlegendentry{GLP, $\lcoef=0.1$}

\addplot [color=mycolor3]
  table[row sep=crcr]{%
0.1	0.85913999\\
0.2	0.83885064\\
0.3	0.82958777\\
0.4	0.86115661\\
0.5	0.85208853\\
0.6	0.82910539\\
0.7	0.84588745\\
0.8	0.84189634\\
};
\addlegendentry{dec., $\lcoef=0$, $\lsig=10$}

\addplot [color=mycolor3, dashed, line width=1.0pt]
  table[row sep=crcr]{%
0.1	1.03728339\\
0.2	1.02840438\\
0.3	1.06785856\\
0.4	1.08160914\\
0.5	1.05729252\\
0.6	1.04317522\\
0.7	1.07496445\\
0.8	1.06357372\\
};
\addlegendentry{dec., $\lcoef=0.1$, $\lsig=10$}

\addplot [color=mycolor4]
  table[row sep=crcr]{%
0.1	0.84658076\\
0.2	0.84932615\\
0.3	0.86205378\\
0.4	0.86500531\\
0.5	0.91505171\\
0.6	0.87751806\\
0.7	0.8719862\\
0.8	0.86280906\\
};
\addlegendentry{dec., $\lcoef=0$, $\lsig=\infty$}

\addplot [color=mycolor4, dashed, line width=1.0pt]
  table[row sep=crcr]{%
0.1	1.06811632\\
0.2	1.03116097\\
0.3	1.06659455\\
0.4	1.07270291\\
0.5	1.06205781\\
0.6	1.07642338\\
0.7	1.07015836\\
0.8	1.08393138\\
};
\addlegendentry{dec., $\lcoef=0.1$, $\lsig=\infty$}

\end{axis}
\end{tikzpicture}%
}%
	\vspace{-2mm}%
	\caption{%
		Elapsed times for declipping on a single frame of a realization of an AR process.
	}
	\label{fig:demo.times}
\end{figure*}

\section{Proximal operators}
\label{sec:Implement.details}
%\todo{\textbf{OM:} Přejmenovat nebo rozdělit na víc částí?}

In this appendix, we treat in more detail the components utilized to fit the proposed regularized AR model in the case of audio declipping,
i.e.,
to solve the problem \eqref{eq:AAR.declipping}.
In \eqref{eq:prox.error}, we already have introduced the proximal operators of the error term appearing in \eqref{eq:AAR.declipping}.
For the sparsity term $\tau\norm{\a}_1$ with any scalar $\tau>0$,
the corresponding proximal operator
%$\prox_{\tau \fcoef} = \prox_{\tau \norm{\cdot}_1}$
$\prox_{\tau \norm{\cdot}_1}$
is the soft thresholding with a~threshold of $\tau$ \cite[Ex.\,6.8]{Beck2017:First.Order.Methods}:
\begin{equation}
	\left(\prox_{\tau \norm{\cdot}_1}(\a)\right)_n
	%= (\abs{a_n}-\lcoef)_+\sgn(a_n)
	= \begin{cases}
		a_n-\tau, & a_n \geq \tau, \\
		0, & \abs{a_n} < \tau, \\
		a_n+\tau, & a_n \leq -\tau.
	\end{cases}
	\label{eq:prox.coef}
\end{equation}
Regarding the signal regularization $\tau d_{\setdec}(\x)^2/2$, we have for $\tau<\infty$
\cite[Ex.\,6.65]{Beck2017:First.Order.Methods}
\begin{equation}
	\prox_{\tau d_{\setdec}(\x)^2/2}(\x) = \tfrac{\tau}{\tau+1}\proj_{\setdec}(\x) + \tfrac{1}{\tau+1}\x.
	\label{eq:prox.sig}
\end{equation}
%
%\todo{\textbf{OM:} Trojitý dolní index, ajaj.\\ PR: Nepřipadne mi to tak zlé :)}%
To cover also the consistent case with $\x\in\setdec$,
we allow the choice $\tau=\infty$,
which is interpreted as replacing $\tau d_{\setdec}(\x)^2/2$ with the indicator function $\iota_{\setdec}(\x)$.
The corresponding proximal operator is the projection \cite[Thm.\,6.24]{Beck2017:First.Order.Methods}
\begin{equation}
	\prox_{\iota_{\setdec}}(\x) = \proj_{\setdec}(\x).
	\label{eq:prox.sig.projection}
\end{equation}
While \eqref{eq:prox.sig.projection} is the limit case of \eqref{eq:prox.sig} for $\tau\to\infty$,
our Matlab implementation treats the two cases individually.

%\begin{itemize}
%	\item \ond{různé implementace DR, šetření paměti apod. (race coef, race signal)?}
%	\item \ond{testování hodnot parametru DR algoritmu (gamma test)?}
%	\todo{\textbf{OM:} Spíš ne, není z toho jasný závěr, viz poznámky výše.}
%	\item \ond{stručná dokumentace přiložených kódů? tady nebo na GitHubu?}
%		\pav{myslím že jen na githubu}
%\end{itemize}


%\ond{%
%\section{}

\section*{Acknowledgment}

The work was supported by the Czech Science Foundation (GAČR) Project 
No.\,23-07294S.
%No.\,20-29009S.\todo{Nový projekt.}

\ifCLASSOPTIONcaptionsoff
  \newpage
\fi



%\newpage

\bibliographystyle{IEEEtran}
% \bibliography{IEEEabrv,../bib/paper}
%\bibliography{../../melodyboys_bibliography/literatura}
\bibliography{literatura}

\end{document}